\documentclass[aps,prd,10pt,nofootinbib,twocolumn,superscriptaddress,floatfix,notitlepage]{revtex4-2}

\usepackage{graphicx}
\usepackage{float}
\usepackage{placeins}

\usepackage{amsmath,amsfonts,amssymb}
\usepackage[dvipsnames]{xcolor}
\usepackage{verbatim}
\usepackage{enumitem}
\usepackage{aas_macros}
\usepackage[breaklinks,colorlinks,urlcolor=MidnightBlue,citecolor=WildStrawberry,linkcolor=Fuchsia]{hyperref}

\newcommand{\Msun}{M_\odot}
\newcommand{\td}{{\rm d}}

\newcommand{\be}{\begin{equation}}
\newcommand{\ee}{\end{equation}}
\newcommand{\bea}{\begin{equation} \begin{aligned}}
\newcommand{\eea}{\end{aligned} \end{equation}}

\def\lsim{\mathrel{\raise.3ex\hbox{$<$\kern-.75em\lower1ex\hbox{$\sim$}}}}
\def\gsim{\mathrel{\raise.3ex\hbox{$>$\kern-.75em\lower1ex\hbox{$\sim$}}}}

\begin{document}

\title{The Loud Tail of the Supermassive Black Hole Binary Population: \texorpdfstring{\\}{ } 
Multimessenger Candidates and Prospects for SKAO}

\author{Juhan Raidal}
\email{juhan.raidal@kbfi.ee}
\affiliation{Laboratory of High Energy and Computational Physics, NICPB, R{\"a}vala 10, Tallinn, 10143, Estonia}
\affiliation{Department of Cybernetics, Tallinn University of Technology, Akadeemia tee 21, 12618 Tallinn, Estonia}

\author{Juan Urrutia}
\email{juan.urrutia@kbfi.ee}
\affiliation{Laboratory of High Energy and Computational Physics, NICPB, R{\"a}vala 10, Tallinn, 10143, Estonia}
\affiliation{Department of Cybernetics, Tallinn University of Technology, Akadeemia tee 21, 12618 Tallinn, Estonia}

\author{Ville Vaskonen}
\email{ville.vaskonen@kbfi.ee}
\affiliation{Laboratory of High Energy and Computational Physics, NICPB, R{\"a}vala 10, Tallinn, 10143, Estonia}

\author{Hardi Veerm\"ae}
\email{hardi.veermae@cern.ch}
\affiliation{Laboratory of High Energy and Computational Physics, NICPB, R{\"a}vala 10, Tallinn, 10143, Estonia}

\begin{abstract}
We assess whether electromagnetically identified supermassive black hole (SMBH) binary candidates populate the rare, high-amplitude tail of the gravitational wave (GW) amplitude distribution. Using an SMBH binary population model fitted to the NANOGrav 15-year data, we find that 3C~66B and Mrk~501 are likely to stand out from the GW background, making them promising targets for individual-source searches. Furthermore, we show that individual detections can probe the binary hardening mechanism, as models with strong environmental hardening predict a higher abundance of individually resolvable binaries than purely GW-driven models. Applying the same population fits and accounting for the confusion noise from the unresolved population, we forecast that the Square Kilometre Array Observatory (SKAO) will resolve tens of individual binaries, with 3C~66B robustly detectable.
\end{abstract}

\maketitle


\section{Introduction}
\label{sec:intro}

Supermassive black holes (SMBHs), with masses in the range $10^6-10^{10}\,\Msun$, reside in the nuclei of nearly all galaxies~\cite{Kormendy:2013dxa}, yet substantial uncertainties persist regarding their formation and their merger rate. If SMBH binaries overcome the final parsec problem and merge efficiently~\cite{Begelman:1980vb,Merritt:2013awa,Kelley:2016gse,Keitaanranta:2025ncl} they will emit ultra-low-frequency gravitational waves (GWs) that add up to a stochastic GW background (SGWB)~\cite{Rajagopal:1994zj,Phinney:2001di,Wyithe:2002ep,Sesana:2004sp} that can be probed with pulsar timing arrays (PTAs)~\cite{Hellings:1983fr,Taylor:2021yjx}. For sufficiently nearby and massive systems, the emission of a single binary could be individually resolvable~\cite{Sesana:2008xk,Rosado:2015epa,Kelley:2017vox,Mingarelli:2017fbe,Mingarelli:2026kjw}. Individual sources are especially valuable when simultaneously identified in the electromagnetic (EM) domain~\cite{DeRosa:2019myq}, as such multimessenger systems provide a direct probe of binary evolution and the mechanisms that drive the binary to sub-parsec separations.

The PTA experiments have reported evidence for an SGWB~\cite{NANOGrav:2023gor,EPTA:2023fyk,Reardon:2023gzh,Xu:2023wog} consistent with an SMBH binary origin~\cite{NANOGrav:2023hfp,EPTA:2023xxk,Ellis:2023dgf}, and the upcoming Square Kilometre Array Observatory (SKAO) will improve the timing sensitivity by more than an order of magnitude~\cite{Janssen:2014dka,SKAOPulsarScienceWorkingGroup:2025oyu}, substantially improving the prospects for individual-source detection~\cite{Sesana:2008xk,Rosado:2015epa}. In parallel, targeted searches have placed upper limits on individual binaries at the sky positions of known EM candidates~\cite{Jenet:2003ew,NANOGrav:2020lwu,NANOGrav:2023wsz,NANOGrav:2025gqp,Tremblay:2025fgk}, a heterogeneous set identified through periodic optical, X-ray, or radio signatures~\cite{DeRosa:2019myq,Graham:2015gma,Severgnini:2018jol,Sudou:2003hv}. Whether any such candidate can plausibly stand out from the SGWB is a statistical question because the loudest sources at a given frequency are drawn from the high-amplitude tail of the same population that produces the background.

The shape of the SGWB spectrum is currently the primary handle on how SMBH binaries evolve. Interactions with stars and gas shorten the residence time at large separations and flatten the spectrum below a few nHz, whereas GW emission alone yields the canonical $f^{-2/3}$ power law~\cite{Sesana:2013wja,Middleton:2015oda,Chen:2016kax,Kelley:2016gse}. With the current data, however, the environmental parameters are degenerate with the overall abundance of binaries since more efficient hardening at low frequencies can be compensated by a larger number of binaries, leaving the spectrum nearly unchanged~\cite{NANOGrav:2023hfp,EPTA:2023xxk,Ellis:2023dgf}. The two scenarios then differ mainly at higher frequencies, where environmental effects are weak, in the number of binaries loud enough to be individually resolved. Individual sources, therefore, carry information about binary evolution that the spectrum alone does not.

In this work, we characterize the SMBH binary population through the GW amplitude distribution (GWAD)~\cite{Raidal:2026ezm}, i.e., the number of binaries per unit GW amplitude at a given observed frequency. The SGWB is set by the bulk of this distribution and the individually resolvable sources by its high-amplitude tail. This provides an efficient framework for mapping out the SGWB as well as the prospects for detecting individual sources in a unified way. Fitting an SMBH binary population model to the NANOGrav 15-year data~\cite{NANOGrav:2023gor}, we identify 3C~66B and Mrk~501 as promising candidates for rare, high-amplitude sources that stand out above the SGWB. Because GW-driven and environmentally driven binary evolution predict different abundances of loud sources, the detection of individual binaries can also probe the binary hardening mechanism, providing information complementary to the SGWB spectrum. Applying the same population model to the projected SKAO sensitivity~\cite{SKAOPulsarScienceWorkingGroup:2025oyu}, we forecast that tens of individual binaries could be resolved once the confusion noise from the unresolved population is accounted for, some of which may become multimessenger sources. Such sources can also serve as standard sirens, enabling measurements of the local expansion history~\cite{Wang:2022oou,Sardana:2026jvr}.

The paper is organized as follows. In Sec.~\ref{sec:gwad} we introduce the GWAD together with the binary evolution and merger rate models, and in Sec.~\ref{sec:pta} we fit them to the NANOGrav 15-year data. Sec.~\ref{sec:MM} confronts the EM candidates with the fitted population, and Sec.~\ref{sec:SKAO} forecasts the population resolvable by SKAO, including the confusion noise from unresolved binaries. We conclude in Sec.~\ref{sec:concl}.

\section{GW amplitude distribution}
\label{sec:gwad}

The GW amplitude distribution (GWAD) describes the distribution of GW amplitudes $A$ emitted by individual binaries at a fixed frequency $f$. It determines the statistical properties of the SMBH GW power spectrum and the associated timing residuals, and for circular binaries, it can be expressed as
\be\label{eq:P1semi-analytic}
    \frac{\td N}{\td A \,\td \ln f} 
    = \int \td\lambda \,\frac{\td t}{\td\ln{f_{\rm b}}}\, \delta \big(A - A^{(1)}\big) \Big|_{f_{\rm b} = \frac{(1+z)f}{2}} \,,
\ee
where $f_{\rm b}$ denotes the rest-frame orbital frequency of the binary related to the observed GW frequency $f$ via $f_{\rm b} = (1+z)f/2$, $\td \lambda$ is the SMBH binary merger rate in the observer frame, $\td t/\td \ln f_{\rm b}$ is the binary residence time, and\footnote{Throughout, we use the geometric units $G=c=1$ and the standard $\Lambda$CDM cosmology with parameters inferred from CMB observations~\cite{Planck:2018vyg}.}
\be \label{eq:A1}
    A^{(1)} \equiv \frac{4 (1+z) \mathcal{M}^{5/3} (2\pi f_{\rm b})^{2/3}}{D_{L}} \,,
\ee
is the GW amplitude from a single circular inspiraling binary with chirp mass $\mathcal{M}=(m_1 m_2)^{3/5}/(m_1+m_2)^{1/5}$ at luminosity distance $D_L$.

The properties of the GWAD are discussed in detail in~\cite{Raidal:2026ezm}. It has a universal heavy power-law tail, $\propto A^{-4}$, while its low-amplitude behavior depends on the SMBH merger rate and the energy-loss mechanisms of the binaries. In the following subsections, we discuss the binary evolution, which enters through $\td t/\td \ln f_{\rm b}$, and introduce the merger rate $\td\lambda$ used in this work. The same merger rate was also used in~\cite{Raidal:2026ezm}. Although the parametrization of the environmental hardening timescale introduced below differs slightly from that used in~\cite{Raidal:2026ezm}, this difference does not affect the general properties of the GWAD.

\subsection{Binary evolution}

The binary orbital evolution can be characterized by the hardening timescale, which describes the rate at which the binary separation $a$ shrinks, $t_{\rm hard}=a/|\dot a|$. Using Kepler’s law, $a\propto M^{1/3}f_{\rm b}^{-2/3}$, the corresponding binary residence time is
\be \label{eq:residence}
    \frac{\td t}{\td \ln f_{\rm b}} = \frac{2}{3}t_{\rm hard} \,.
\ee
At sufficiently high GW frequencies, energy loss through GW emission is expected to dominate the evolution. For a circular binary, the corresponding hardening timescale is~\cite{Peters:1963ux},\footnote{A non-zero eccentricity enhances the GW emission and redistributes it over harmonics of the orbital frequency~\cite{Peters:1963ux}. A hard binary formed with high eccentricity merges much faster than a circular one.}
\be \label{eq:tcirc}
    t_{\rm hard} \approx t_{\rm GW} = \frac{5}{64} \frac{1+z}{\mathcal{M}^{5/3} (2 \pi f_{\rm b})^{8/3}} \,,
\ee 
At lower frequencies, the evolution can instead be dominated by interactions with ambient stars and gas~\cite{Armitage:2002uu,2013CQGra..30x4005M,Kelley:2016gse,Tang:2017eiz}. We characterize these environmental effects by a timescale $t_{\rm env}$, which modifies the hardening timescale as 
\be
    t_{\rm hard} = \frac{1}{t_{\rm GW}^{-1} + t_{\rm env}^{-1}} \,.
\ee
We parameterize the hardening timescale through environmental effects as
\be \label{eq:tenv}
    t_{\rm env} = t_{\rm ref} \, (1+z) (4\eta)^\gamma  \left(\frac{f_{\rm b}}{\rm nHz}\right)^\alpha \left(\frac{M}{10^9 \,\Msun}\right)^\beta \,,
\ee
where $M = m_1 + m_2$ and $\eta=m_1 m_2/(m_1+m_2)^2$, such that $4\eta\in(0,1]$. This parametrization differs slightly from that adopted in our previous works~\cite{Ellis:2023dgf}, with the addition of the parameter $\gamma$, which controls the mass-ratio dependence.\footnote{Eq.~\eqref{eq:tenv} extends the parametrization used in~\cite{Ellis:2023dgf,Raidal:2026ezm}, which we denote as $(\alpha',\beta')$, and relates to the current one as $\alpha=\alpha'-8/3$, $\beta=\alpha'\beta'-5/3$, $\gamma = 3\beta/5$.} Related environmental models have been considered in Refs.~\cite{2009ApJ...700.1952H,Sesana:2013wja,Kelley:2016gse,Kozhikkal:2023gkt,NANOGrav:2023hfp}. To determine the range of values that the parameters entering $t_{\rm env}$ can take, we briefly summarize the different mechanisms that can drive energy loss from the binary.

At large separations, before a bound binary forms, dynamical friction against the stellar background~\cite{Chandrasekhar:1943ys} causes the separation between the SMBHs to shrink on a timescale $t_{\rm df} \propto a^2\sigma/m_2$, where $\sigma$ is the stellar velocity dispersion. Once a bound binary has formed, stars entering the binary’s loss cone undergo three-body interactions and are ejected, extracting energy and angular momentum from the binary and thereby hardening it~\cite{Begelman:1980vb,Quinlan:1996vp}. The corresponding hardening timescale scales as $t_{\rm sh} \propto \sigma/(\rho a)$, where $\rho$ and $\sigma$ denote the stellar mass density and velocity dispersion in the vicinity of the binary. 

If the binary happens to be formed in a gas-rich nucleus, the binary couples to a circumbinary disk with supply rate $\dot M_{\rm d} = 3\pi\nu\Sigma$, where $\nu$ denotes the kinematic viscosity and $\Sigma$ the surface mass density of the disc. The radial gas velocity is $v_r=\dot M_{\rm d}/(2\pi\Sigma a)$~\cite{Haiman:2009te}. In the disc-dominated case, $m_2\lesssim\Sigma a^{2}$, the secondary BH is carried inward with the gas, $\dot a\simeq -v_r \propto -\nu/a$, corresponding to a hardening timescale $t_{\rm disc} \propto a^{2}/\nu$. If the secondary is instead heavier than the local disc, $m_2\gtrsim\Sigma a^{2}$, its inertia limits the migration rate. The available viscous mass flux is insufficient to carry the secondary inward at the gas drift velocity, suppressing the migration rate by the ratio of the local disk and secondary masses, which gives a hardening timescale $t_{\rm sec}\propto m_2/(\nu\Sigma)$~\cite{Syer:1995hk,Ivanov:1998qk}. Since $\nu\Sigma=\dot M_{\rm d}/3\pi$, this timescale can equivalently be written as $t_{\rm sec}\propto m_2/\dot M_{\rm d}$. For a steady disk fed at a constant rate, $\dot M_{\rm d}$ is independent of separation, and the hardening timescale is therefore independent of frequency~\cite{Kocsis:2010xa}.

\begin{table}
\centering
\renewcommand{\arraystretch}{1.0}
\setlength{\tabcolsep}{3pt}
\begin{tabular}{lccc}
\hline
Mechanism & $\alpha$ & $\beta$ & $\gamma$ \\
\hline
Dynamical friction            & $-4/3$ & $(-1/3,\,-2/15)$ & $-1$ \\
Stellar loss-cone scattering                & $2/3$  & $(-1/3,\,-2/15)$ & $0$  \\
Viscous drag (disc-dom.)      & $-1$   & $(-1/2,\,1/2)$   & $0$  \\
Viscous drag (sec.-dom.)          & $-1/3$ & $(1/6,\,2/3)$    & $1$         \\
Gas infall/accretion (sec.-dom.)  & $0$    & $(0,\,1)$        & $1$         \\
\hline
GW emission     & $-8/3$ & $-5/3$           & $-1$        \\
\hline
\end{tabular}
\caption{The parameter values of the environmental hardening timescale parametrization~\eqref{eq:tenv} for the mechanisms discussed in the text.}
\label{tab:env}
\end{table}

In Table~\ref{tab:env}, we give the exponents of $t_{\rm env}$ obtained from the hardening timescales discussed above, using Kepler’s law and the disk viscosity $\nu\propto(H/r)^{2}M^{1/2}a^{1/2}$~\cite{Shakura:1972te}. For the stellar effects, the $\beta$ interval spans from the constant-environment value $\beta=-1/3$ to the value $\beta=-2/15$ obtained for $\sigma\propto M^{1/5}$~\cite{Kormendy:2013dxa}. Similarly, for secondary-dominated viscous drag, $\Sigma\propto 1/a$ gives $\beta=2/3$, while $\Sigma\propto M^{1/2}/a$~\cite{Kocsis:2012ui} gives $\beta=1/6$. The disk-dominated interval is instead centered on the constant-environment value $\beta=0$, with its endpoints set by $H/r\propto M^{\delta}$, where $\delta\in(-1/4,1/4)$ spans across the gas- and radiation-pressure-dominated regions of the disc~\cite{Kocsis:2012ui}. For gas infall, we fix $\dot M_{\rm d}$ directly, and the interval extends from the Eddington-limited case, $\dot M_{\rm d}\propto M$, to the mass-independent case.

\subsection{Merger rate}

The SMBH binary population is described by its comoving merger rate density, which enters the merger rate in the observer frame as
\be \label{eq:diffmergerrate}
    \td \lambda = \td \mathcal{M} \td \eta \td z \, \frac{1}{1+z} \frac{\td V_c}{\td z} \frac{\td R_{\rm BH}}{\td \mathcal{M} \td \eta}\,,
\ee
where the factor $1/(1+z)$ converts the source-frame rate to the observer frame and the derivative of the comoving volume is
\be
    \frac{\td V_c}{\td z} = \frac{4\pi}{H} \frac{D_L^2}{(1+z)^2} \,,
\ee
in terms of the Hubble rate $H$ and luminosity distance $D_L$. The merger rate density per unit chirp mass and symmetric mass ratio, $\td R_{\rm BH}/\td\mathcal{M}\,\td\eta$, is obtained from the rate per unit component mass $\td R_{\rm BH}/\td m_1\,\td m_2$ by the corresponding change of variables. The latter follows from the halo merger rate $R_h$ as
\be \label{eq:modelA}
    \!\!\frac{\td R_{\rm BH}}{\td m_1 \td m_2} 
    \!=\! p_{\rm BH} \!\!\int \!\!\td M_1 \td M_2 \frac{\td R_h}{\td M_1 \td M_2} \!\!\prod_{j=1,2} \!\!\frac{\td P(m_j|M_j)}{\td m_j}\!,
\ee
where $m_j$ are the SMBH masses, $M_j$ their host halo masses, and $p_{\rm BH} \le 1$ encodes both the SMBH occupation fraction and the probability that the black holes merge following their host halos. We use the halo merger rate from the extended Press--Schechter formalism~\citep{Press:1973iz,Bond:1990iw,1993MNRAS.262..627L}. Halo masses are mapped to galaxy stellar masses $M_*(M_j)$ using Ref.~\cite{Girelli:2020goz}, and the BH--stellar mass relation is
\be \label{eq:Ms_relation}
    \frac{\td P(m|M_*)}{\td \log_{10} m} = \mathcal{N}\bigg(\log_{10} \frac{m}{\Msun} \bigg| a + b \log_{10} \frac{M_*}{10^{11}\Msun},\sigma\bigg) \,,
\ee
where $\mathcal{N}(x|\bar x,\sigma)$ is a Gaussian with mean $\bar x$ and variance $\sigma^2$. Fits to dynamically measured local SMBHs give $a = 8.95$, $b = 1.4$, and $\sigma = 0.47$~\cite{Reines:2015nyy}, which we adopt as fiducial values.

\begin{figure*}
    \centering
    \includegraphics[width=\linewidth]{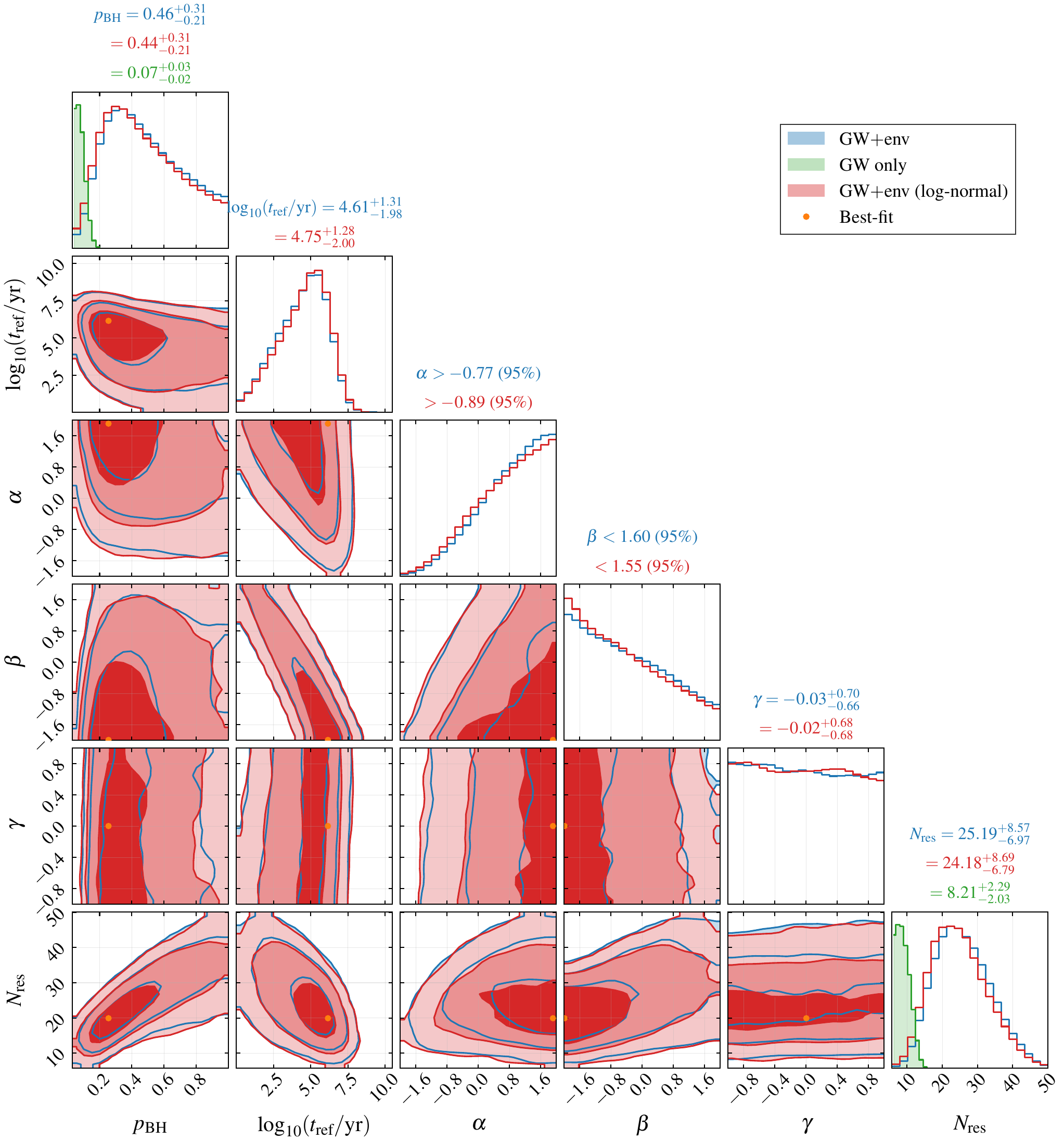}
    \caption{Posteriors of the SMBH binary model fit against the NANOGrav 15-year data. The contours indicate the $1,2,3\sigma$ credible regions. We also show the posteriors of the expected number of SKAO-resolvable binaries, $N_{\rm res}$, derived from the posteriors of model parameters.}
    \label{fig:corner}
\end{figure*}

\section{PTA fit}
\label{sec:pta}

PTA analyses model the GW-induced timing residuals as a Gaussian stochastic process whose power spectrum is either fitted to a parametric spectral model or estimated independently in each frequency bin. To estimate the PTA likelihood for SMBH-binary population inference, we employ a variance-averaged Gaussian approximation~\cite{Ellis:2023dgf,Xue:2024qtx,Raidal:2026ezm}. This approximation accounts for the full non-Gaussian distribution\footnote{Here, non-Gaussianity refers to the realisation-to-realisation scatter of the timing residual variance $\sigma_k^2$, which is heavy-tailed and enters through the likelihood~\eqref{eq:likelihood}. It is distinct from the non-Gaussianity of the timing residuals across pulsars within a single realisation and frequency bin, which has been shown to be practically inaccessible to model-agnostic tests~\cite{Cecchini:2026cru}.} of the timing-residual variance after averaging over binary phases, polarizations, inclinations, and sky locations~\cite{Raidal:2026ezm},
\be \label{eq:OmegaGW}
	\sigma_k^2 \equiv \langle |\tilde{\delta t}_{k}|^2 \rangle_{\delta, \psi, \imath, \hat{k}} = \frac{1}{60 \pi^2} \sum_{j=1}^{N(f_k)} \frac{A_j^2}{f_j^2} \,,
\ee
where $k$ labels the PTA Fourier frequencies and the sum runs over the $N(f_k)$ binaries emitting in the $k$th bin. The likelihood combines the predicted distribution $\td P/\td\sigma_k$ with the free-spectrum posteriors $p_k(\sigma_k)$ from the Gaussian-process PTA analysis
\be \label{eq:likelihood}
    \mathcal{L}(\vec{\theta}) = \prod_k \int \td\sigma_k^2 \frac{\td P(\sigma_k^2|f_k,\vec{\theta})} {\td\sigma_k^2} p_k(\sigma_k^2) \,,
\ee 
where $\vec\theta=\{p_{\rm BH},t_{\rm ref},\alpha,\beta,\gamma\}$ collects the population and environmental parameters.  We use this approximation with the \texttt{GWADpy} code~\cite{Raidal_GWAD_2026}.

We fit to the NANOGrav 15-year Hellings–Downs correlated free-spectrum posteriors. Rather than the originally published free spectrum~\cite{NANOGrav:2023gor}, we use the reanalysis of~\citet{Agarwal:2026dlv}, which replaces the DMX treatment of dispersion-measure variations with the customized chromatic-noise models (CNMs) of~\citet{Larsen:2026pru}. The most consequential change for our purposes is at 15.8 nHz, where the published free spectrum showed an excursion above the power-law expectation that the reanalysis attributes to non-stationary chromatic transients in PSR J1713+0747. The recovered amplitude in that bin falls by 0.27 dex and its posterior broadens substantially. The spectrum is also slightly quieter overall, with the power-law amplitude at fixed $\gamma = 13/3$ being reduced by 12\%, and slightly steeper, $\gamma = 3.5^{+0.7}_{-0.6}$ against the originally reported $\gamma = 3.2^{+0.6}_{-0.6}$.

Although the data spans $30$ bins over $1.98$--$59.3\,$nHz, we retain the lowest $14$ bins, above which the posteriors become uninformative and consistent with zero. We perform a five-parameter fit for the binary formation efficiency $p_{\rm BH}$ and the environmental timescale parameters $\{t_{\rm ref},\alpha,\beta,\gamma\}$ of Eq.~\eqref{eq:tenv}, which enter through the GWAD. The MCMC is performed using the Python \texttt{emcee} package, adopting uniform priors
\bea
    &p_{\rm BH} \in [0,\,1] \,, \qquad \log_{10}(t_{\rm ref}/{\rm yr}) \in [0,\,12] \,, \\
    &\alpha \in [-2,\,2] \,, \quad \beta \in [-2,\,2] \,, \quad \gamma \in [-1,\,1] \,,
\eea
motivated by the mechanisms listed in Table~\ref{tab:env}.

For each proposal, the GWAD of Eq.~\eqref{eq:P1semi-analytic} is evaluated, and the distribution of the timing residual variance is built from $10^5$ Monte-Carlo realizations of the binary population, with an expected $50$ strong sources per source frequency bin and $200$ logarithmic frequency bins covering the frequency range of the NG15 data. The amplitudes of the strong binaries are drawn explicitly from GWAD, while the remainder are accumulated into a Gaussian background. Chain convergence is reached when the Gelman-Rubin statistic $R-1<0.01$ for each parameter and the autocorrelation times, $\tau$, for all samples, $N$, yield $N/\tau>50$.

The resulting posteriors are shown in Fig.~\ref{fig:corner}. The binary formation efficiency exhibits a lower bound, $p_{\rm BH} > 0.17$ at $95\%$ confidence, with a mild preference for values around $p_{\rm BH}\approx 0.5$. The reference timescale $t_{\rm ref}$ instead has a broad maximum near $\log_{10}(t_{\rm ref}/{\rm yr})\simeq 5\text{--}6$ and a long tail toward shorter timescales. The GW-driven limit corresponds to the opposite, large $t_{\rm ref}$ regime, where the preferred binary formation efficiency is significantly lower, $p_{\rm BH}\simeq 0.07$, as shown by the green posteriors. 

Compared to the fit with the original DMX free spectra~\cite{NANOGrav:2023gor}, the CNM free spectrum lowers the inferred binary formation efficiency by about a third, from $p_{\rm BH} = 0.69^{+0.21}_{-0.26}$ to $p_{\rm BH} = 0.46^{+0.31}_{-0.21}$, while leaving the environmental parameters statistically unchanged. Since the re-calibrated data is more consistent with a $\gamma=13/3$ power law in~\citet{Agarwal:2026dlv}, the preference for environmental hardening over purely GW-driven evolution in our analysis weakens accordingly, from $\Delta \mathcal{L} = 9.4$ ($\Delta {\rm BIC} = 8.2$) to $\Delta \mathcal{L} = 7.9$ ($\Delta {\rm BIC} = 5.2$), but is not removed.

The reference timescale is degenerate with the exponent $\alpha$, which determines the frequency dependence of the environmental hardening timescale. Larger values of $\alpha$ imply a stronger increase of $t_{\rm env}$ with frequency and are therefore compensated by smaller values of $t_{\rm ref}$. The data provide a lower bound $\alpha > -0.77$, with the posterior rising with $\alpha$. Strongly negative values, corresponding to dynamical-friction- or disk-dominated viscous-drag hardening, are therefore disfavored, whereas shallower scalings, such as those associated with stellar loss-cone scattering or gas infall, remain fully consistent with the data. The mass exponent shows the opposite behavior, with only an upper limit $\beta < 1.60$ and a marginal posterior that grows towards the lower prior edge. The mass-ratio exponent $\gamma$ has a flat posterior and remains largely unconstrained.

The corner plot further compares the posteriors obtained with the variance-averaged and purely log-normal treatments. In the latter, the distribution $\td P/\td \sigma_k^2$ is approximated by a log-normal distribution whose mean and variance match those of the actual distribution. We find that the two treatments yield near-identical posteriors, indicating that the population-level inference of the NANOGrav collaboration~\cite{NANOGrav:2023hfp} remains adequate despite the heavy-tailed statistics governing the individual-source regime of the GWAD.

\section{Multimessenger candidates}
\label{sec:MM}

SMBH binary candidates have been identified through periodic EM signatures~\cite{DeRosa:2019myq}. Those best suited for multimessenger studies are nearby high-mass systems emitting in the PTA band. We have collected seven such candidates in Table~\ref{tab:candidates}. OJ~287 shows double-peaked optical outbursts every 12\,yr, attributed to a secondary on an eccentric ($e\simeq0.65$) orbit crossing the disk of a $\mathcal{O}(10^{10}\Msun)$ primary twice per orbit~\cite{1996ApJ...460..207L,Laine:2020dnr}. Mrk~501 has been proposed as a binary candidate based on periodic changes in its jet morphology, interpreted as evidence for orbital motion~\cite{Ghisellini:2009fj,Britzen:2026stag}.\footnote{Non-binary explanations for the jet features of Mrk~501, including jet instabilities, frequency-dependent core shifts, and internal shocks, were found to be disfavored in Ref.~\cite{Britzen:2026stag}.} Other candidates have been identified through optical periodicity (PG~1302-102), X-ray variability and iron K$\alpha$ profiles (MCG+11-11-032), or AGN variability (Mrk~915 and J~1536+0441)~\cite{Graham:2015gma,Severgnini:2018jol,Serafinelli:2020nog}. The clearest kinematic evidence comes from 3C~66B, whose radio core shows VLBI-resolved elliptical motion with a $1.05$\,yr period~\cite{Sudou:2003hv,2010ApJ...724L.166I}. Some of these systems have been targeted in dedicated PTA searches, with 3C~66B receiving the most extensive attention~\cite{Jenet:2003ew,NANOGrav:2020lwu,NANOGrav:2023wsz,NANOGrav:2025gqp,Tremblay:2025fgk}. The most recent study \cite{Tremblay:2025fgk} placed 95\% upper limits on the chirp mass of 3C 66B from Parkes PTA data. Under an astrophysically motivated log-uniform mass prior they obtain $\mathcal{M}<6.9\times10^8\Msun$, while under a prior uniform in the mass — matching the one we adopt here — the limit relaxes to $\mathcal{M}<10.63\times10^8\Msun$, excluding only the upper 2\% of our mass prior. The corresponding amplitude limit is $A < 1.3\times10^{-14}$. 

\begin{figure*}
    \centering
    \includegraphics[width=0.98\linewidth]{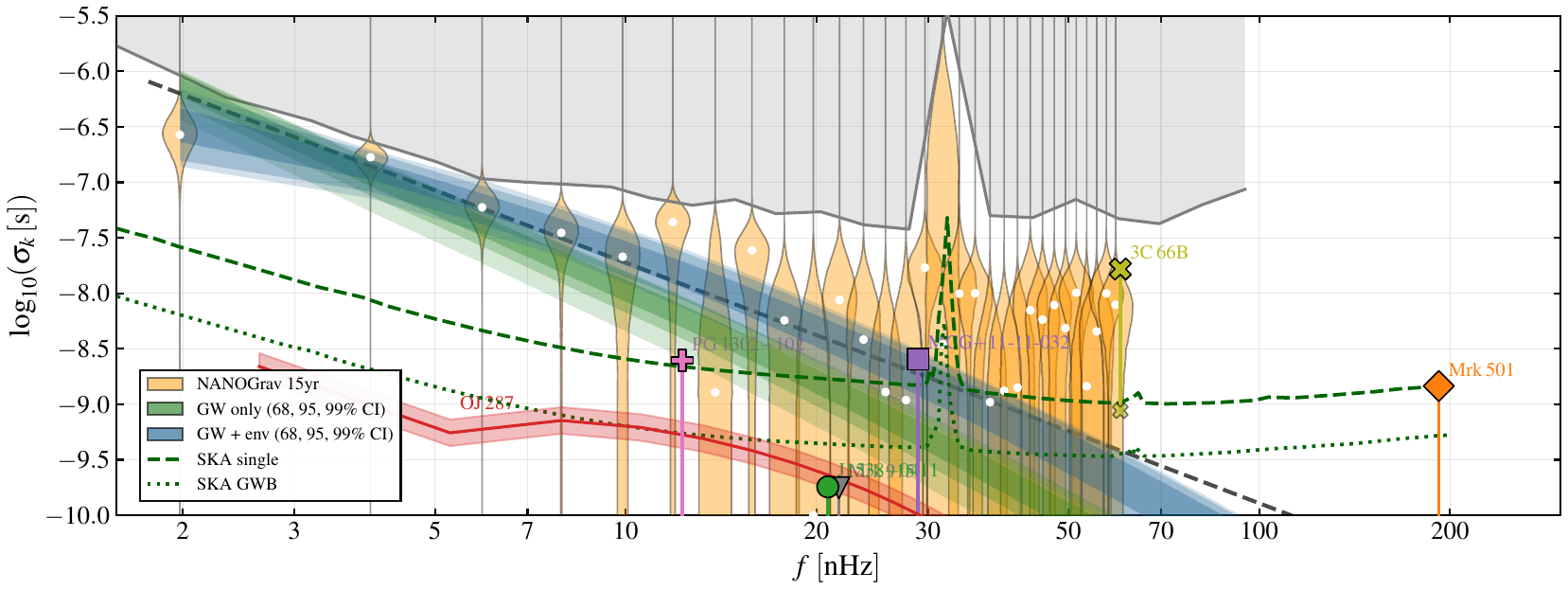}
    \caption{Electromagnetically identified SMBH binary candidates (Table~\ref{tab:candidates}) in the plane of characteristic timing residual $\sigma_k$ versus GW frequency, placed at $f_{\rm GW}=2/T$ with the exception of OJ287, which is denoted by the red band. The uncertainty in mass ratio is shown by a solid line connecting the upper and lower magnitude estimates. The orange violins show the NANOGrav 15-year free-spectrum posteriors under the customized chromatic-noise models \cite{Agarwal:2026dlv, Larsen:2026pru}, and the shaded bands show the $68\%$, $95\%$, and $99\%$ ranges of the distribution obtained by mapping the posteriors shown in Fig.~\ref{fig:corner} onto the median GW spectrum in the GW-only (green) and GW+environment (blue) cases. The dashed and dotted green curves show the projected single-source and stochastic-background sensitivities of SKAO. The gray shading shows NANOGrav 15-year bound for individual SMBH binaries~\cite{NANOGrav:2023pdq}.}
    \label{fig:candidates}
\end{figure*}

\begin{table*}
\centering
\begin{tabular*}{\textwidth}{@{\extracolsep{\fill}}lccccc|cccc@{}}
\hline
Source & $z$ & $m_1+m_2$ & $m_2$ & $e$ & $T$ & $A$ & $f_{\rm GW}$ & $t_{\rm GW}$ & $\dot f_{\rm GW}T_{\rm obs}^2$ \\
 & & $[\Msun]$ & $[\Msun]$ & & $[{\rm yr}]$ & $[10^{-16}]$ & $[{\rm nHz}]$ & $[{\rm kyr}]$ & \\
\hline
3C~66B~\cite{Sudou:2003hv,2010ApJ...724L.166I} & $0.021$ & $(1.9_{-0.7}^{+0.7})\times 10^9$ & $(7.0_{-6.4}^{+4.7})\times 10^8$ & ? & $1.05$ & $33-260$ & $60.4$ & $2.1$ & $0.55$ \\
Mrk~501~\cite{Ghisellini:2009fj,Britzen:2026stag} & 0.0336 & $(1-10)\times 10^8$ & ? & ? & $0.33$ & $1.5-68$ & $192$ & $0.80$ & $4.5$ \\
MCG+11-11-032~\cite{Severgnini:2018jol,Serafinelli:2020nog} & 0.0362 & $(5.0_{-2.5}^{+5.0})\times10^{8}$ & ? & ? & $2.19\pm 0.05$ & $1.8-18$ & $28.9$ & $120$ & $4.4\times10^{-3}$ \\
PG~1302-102~\cite{Graham:2015gma} & 0.2784 & $(0.2-2.5)\times10^{9}$ & ? & ? & $5.16\pm 0.24$ & $0.11-7.4$ & $12.3$ & $160$ & $1.4\times10^{-3}$ \\
OJ~287~\cite{1996ApJ...460..207L,Laine:2020dnr} & 0.306 & $1.85\times10^{10}$ & $1.5\times10^{8}$ & 0.65 & $12$ & $1.2$ & $21.1$ & $37$ & $1.1\times10^{-2}$ \\
Mrk~915~\cite{Serafinelli:2020nog} & 0.0241 & $(1.1_{-0.4}^{+0.4})\times10^{8}$ & ? & ? & $2.92\pm 0.17$ & $0.26-0.93$ & $21.7$ & $3.4\times10^{3}$ & $1.2\times10^{-4}$ \\
J1536+0441~\cite{Shen:2010aa,Graham:2015tba} & 0.379 & $6.61\times10^{8}$ & ? & ? & $3.04$ & $0.92$ & $20.9$ & $120$ & $3.3\times10^{-3}$ \\
\hline
\end{tabular*}
\caption{Candidate SMBH binaries. The GW amplitude $A$ [Eq.~\eqref{eq:A1}] and GW-driven inspiral timescale $t_{\rm GW}$ [Eq.~\eqref{eq:tcirc}] are evaluated using the measured masses for OJ~287 and 3C~66B, while for the remaining sources, for which $m_2$ is unconstrained, we assume an equal-mass binary ($q=1$). The quoted ranges in $A$ propagate the mass uncertainties at fixed $q$, while the gravitational hardening timescale $t_{\rm GW}$ and the frequency drift $\dot f_{\rm GW}T_{\rm obs}^2$ (with $T_{\rm obs} = 20 \, \rm yr$) are quoted at the central mass, taken to be the midpoint of the quoted interval for the sources with only a mass range available. All sources are treated as circular except OJ~287, for which the quoted frequency $f_{\rm GW}=8/T$ and amplitude correspond to the peak $ n=8$ harmonic of its eccentric ($e=0.65$) orbit, and whose $t_{\rm GW}$ includes the eccentricity enhancement of the GW luminosity.}
\label{tab:candidates}
\end{table*}

Table~\ref{tab:candidates} lists the redshift, masses, eccentricity, and observed period $T$ of each system, together with the derived GW amplitude $A$, frequency $f_{\rm GW}$, and inspiral timescale $t_{\rm GW}$. For the eccentric OJ~287 the quoted dominant $n=8$ mode at $21.1\,$nHz carries $\mathcal{O}(10\%)$ of the power. Treating the other sources as circular, all GW power is emitted at $f_{\rm GW}=2/T$, spanning $\sim 12 - 190\,$nHz and thus the upper PTA band. Only OJ~287 and 3C~66B have both masses measured. For the other systems, we compute $A$ and $t_{\rm GW}$ assuming an equal-mass binary. 

Although $t_{\rm GW}$ exceeds the observation time $T_{\rm obs} = \mathcal{O}(10\,{\rm yr})$ for every candidate, this alone does not guarantee that the source appears monochromatic. Since a PTA resolves frequencies only to $\Delta f \simeq 1/T_{\rm obs}$, the relevant condition is that the frequency drift over the campaign stays within a single Fourier bin, $\dot f_{\rm GW}T_{\rm obs}^2 \lesssim 1$. Combining Eqs.~\eqref{eq:residence} and~\eqref{eq:tcirc}, we can see that $\dot f_{\rm GW}T_{\rm obs}^2 = \tfrac{3}{2}\,(f_{\rm GW}T_{\rm obs})\,(T_{\rm obs}/t_{\rm GW})$, so it corresponds to the naive ratio $T_{\rm obs}/t_{\rm GW}$ enhanced by the large number of cycles $f_{\rm GW}T_{\rm obs} \sim 10-10^2$ accumulated over the observation. This condition is therefore a considerably stronger requirement than $t_{\rm GW} \gg T_{\rm obs}$. 

Evaluating for $T_{\rm obs}=20\,$yr, the five lowest-frequency candidates give $\dot f_{\rm GW}T_{\rm obs}^2 \lesssim  10^{-2}$ and are safely monochromatic. The two loudest candidates do not: 3C~66B is marginal, with $\dot f_{\rm GW}T_{\rm obs}^2 \simeq 0.5$ at its central mass and spanning $0.05-1$ across its mass range. Mrk~501 lies firmly in the evolving regime with $\dot f_{\rm GW}T_{\rm obs}^2 \simeq 5$, reaching $\sim 10$ at the top of its mass range. For these two systems, a monochromatic template loses coherence over $T_{\rm obs}=20\,$yr, and an evolving waveform is required. Moreover, their short inspiral times, $t_{\rm GW}\sim\,\rm kyr$, are comparable to the light-travel delay $d_{\rm p}/c \simeq 3\,\rm kyr$ to a pulsar at $d_{\rm p}\simeq1\,\rm kpc$, so the pulsar term samples the binary at a frequency a factor of a few below the Earth term and cannot be assumed to be coherent with it. Both effects change the waveform model rather than the achievable signal-to-noise ratio and are therefore not obstacles to detection.

\begin{figure*}
    \centering
    \includegraphics[width=0.98\linewidth]{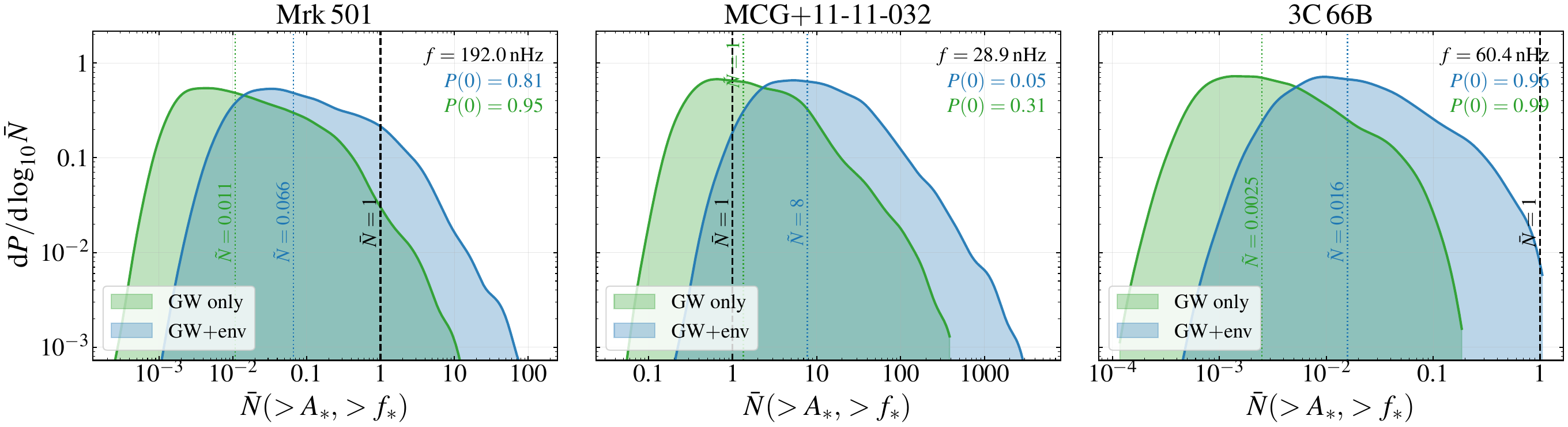}
    \caption{Posterior distributions of the expected number of binaries that are louder and at higher frequency than the candidate binaries in the GW-only (green) and GW+environment (coloured) models. Vertical dotted lines mark the medians and the dashed black line indicates $\bar N=1$. The probability of absence of louder sources, $P(0)$, is quoted for each model in the panel titles.}
    \label{fig:expected}
\end{figure*}

Fig.~\ref{fig:candidates} places the candidates in the plane of characteristic timing residual $\sigma_k$ versus GW frequency. We estimate the characteristic timing residuals induced by the binary candidates by the sky position, polarization and inclination average~\cite{Raidal:2026ezm}, $\sigma_k = A/(\sqrt{60} \pi f_{\rm GW})$, making it directly comparable to the posteriors of the Gaussian-process PTA analysis. The candidates are shown over the NANOGrav 15-year free-spectrum posteriors and distribution of the median GW spectra of our fitted population, together with the projected SKAO single-source and SGWB sensitivities~\cite{SKAOPulsarScienceWorkingGroup:2025oyu}. Most candidates fall near or below the bulk of the fitted spectra, but the nearby high-mass Mrk~501 and 3C~66B lie in the high-amplitude tail of the GWAD, with 3C~66B rising above the SKAO single-source curve.

To quantify whether these candidates occupy the rare high-amplitude tail, Fig.~\ref{fig:expected} shows, for the most promising candidates, the posterior of the expected number of binaries that are both louder $(A>A_{*})$ and at a higher frequency $(f>f_{*})$ than the candidate, obtained by integrating the GWAD:
\be\label{eq:barN}
    \bar N(>A_{*},>f_{*}) = \int_{f_{*}} \!\td \ln f \int_{A_{*}} \!\td A \, \frac{\td N}{\td A \,\td \ln f}\,.
\ee
These posteriors are obtained by mapping the PTA fit posteriors of Fig.~\ref{fig:corner} to $\bar{N}$ and marginalizing over the candidate mass ratio $q = m_2/m_1 \in [0.01,1]$ and the allowed range of its total mass, assuming a circular orbit at the median observed period $T$. A candidate is a rare outlier when $\bar N \lesssim 1$, while $\bar N \gg 1$ indicates that many louder population sources are expected. 

The probability that the candidate is the loudest source at $f > f_{*}$ is obtained by integrating the Poisson probability of finding no louder source over the posterior of $\bar N$:
\be
    P(0) = \int \td \bar{N} \, \frac{\td P}{\td \bar{N}} e^{-\bar{N}} \,.
\ee
For GW-driven evolution, this gives $P(0) = 0.95,\, 0.31,\, 0.99$ for Mrk~501, MCG+11-11-032, and 3C~66B, respectively, while the model with environmental effects gives $P(0) = 0.81,\, 0.05,\, 0.96$. Hence, 3C~66B and Mrk~501 are rare outliers above the SGWB, whereas MCG+11-11-032 is consistent with being a typical member of the background population. 

Being a rare outlier ($\bar N\lesssim1$) also bears on individual resolvability, since events from a crowded part of the population ($\bar N\gg1$) can be difficult to separate from each other even when each exceeds the detection threshold. The relevant criterion is that the expected number of comparable events within the resolvable volume of parameter space is below unity, and $\bar N$ of Eq.~\eqref{eq:barN} is a conservative estimate of this number for two reasons. First, Eq.~\eqref{eq:barN} counts all sources with $A>A_*$ and $f>f_*$, whereas only those within the frequency resolution $f_*\pm1/(2T_{\rm obs})$ can be confused with the candidate, and their expected number is much smaller than $\bar N$. Second, if the GW source is localized to a fraction $f_{\rm sky}$ of the sky, the expected number of comparable events is reduced further to $f_{\rm sky}\bar N$, which can turn a typical event ($\bar N\gg1$) into a rare one ($f_{\rm sky}\bar N\lesssim1$) when $f_{\rm sky}\ll1$. Similarly, a good multimessenger candidate should occupy a region of parameter space where the expected number of GW-resolvable binaries is comparable to unity or smaller. We return to this in Sec.~\ref{sec:SKAO}.

The two hardening models imply systematically different posterior for $\bar{N}$. As indicated by Eq.~\eqref{eq:tenv}, environmental coupling shortens the residence time at large separations, accelerating binaries through the low-frequency regime and depleting the population there. Consequently, as shown in Fig.~\ref{fig:corner}, the PTA fit including environmental effects prefers a markedly higher binary formation probability, $p_{\rm BH}=0.46^{+0.31}_{-0.21}$, compared to purely GW-driven evolution for which $p_{\rm BH}=0.07^{+0.03}_{-0.02}$. This enhanced formation probability increases the abundance of strong sources at the higher frequencies where environmental effects are weak, as reflected in the larger $\bar{N}$ of the environmental model in Fig.~\ref{fig:expected}. Loud sources such as 3C~66B are therefore better compatible with the model where environmental effects are strong than with purely GW-driven evolution. A confirmed individual detection of such a source, either through EM observations or GWs, would thus provide a discriminator between the different hardening scenarios, complementary to the information encoded in the shape of the SGWB spectrum.

\section{SKAO prospects}
\label{sec:SKAO}

The order-of-magnitude sensitivity gain of SKAO~\cite{SKAOPulsarScienceWorkingGroup:2025oyu} will bring individual SMBH binaries within reach. The detectability, however, will become limited over a large part of the frequency range by the confusion noise arising from the SMBH background. We integrate confusion noise into our resolvable sources estimate through an iterative process akin to \cite{SKAOPulsarScienceWorkingGroup:2025oyu}.

The signal-to-noise ratio (SNR) of a monochromatic signal with polarization amplitude $A^{(1)}$ at frequency $f$, averaged over sky position, polarization, and inclination, is
\be \label{eq:rho}
    \rho = \frac{A^{(1)}}{A_{\rm min}(f)} \,,
\ee
where the threshold amplitude $A_{\rm min}(f)$ is obtained from the correspondingly averaged SKAO single-source detection sensitivity curve $h_{\rm c}^{\rho=1}(f)$ given in~\cite{SKAOPulsarScienceWorkingGroup:2025oyu} for observation time $T_{\rm obs}=20\,$yr and number of pulsars $N_p = 174$. It is defined by $\rho = 1$~\cite{Hazboun:2019vhv}, so
\be \label{eq:Amin}
    A_{\rm min}(f) = \frac{h_{\rm c}^{\rho=1}(f)}{\sqrt{f T_{\rm obs}}} \,.
\ee
This does not yet include the noise from unresolved binaries, which degrades the effective sensitivity threshold from $A_{\rm min}$ to $A_{\rm min}^{\rm eff}$, estimated next. 

An individually resolvable source is defined as one whose SNR exceeds a threshold $\rho_{\rm th}$,
\be \label{eq:resolvable}
    A^{(1)} > \rho_{\rm th} A_{\rm min}^{\rm eff}(f) \,,
\ee
for which we set $\rho_{\rm th}=3$ following~\cite{SKAOPulsarScienceWorkingGroup:2025oyu}. Accordingly, the unresolved binaries add a confusion background whose characteristic strain we approximate by the ensemble average obtained by integrating the GWAD, weighted by the energy density emitted by a single binary, over the unresolved population,
\be \label{eq:hcGW}
    h_{\rm c}^{\rm GW}(f)^2 \approx \frac{2}{5} \int_0^{\rho_{\rm th} A_{\rm min}^{\rm eff}(f)} \! {\rm d} A \, A^2 \frac{{\rm d} N}{{\rm d} A \, {\rm d} \ln f} \,.
\ee
The ensemble average in Eq.~\eqref{eq:hcGW} is adequate because the upper limit $\rho_{\rm th}A_{\rm min}^{\rm eff}$ removes the loud tail of the GWAD, so that the confusion noise is dominated by the numerous faint binaries. The factor $2/5$ converts the polarization amplitude into the mean-square strain of a binary. This factor does not appears in Eq.~\eqref{eq:Amin} because it is already included in the definition of $h_{\rm c}^{\rho=1}$~\cite{Hazboun:2019vhv}.

The unresolved binaries act as an additional noise component in each pulsar's timing residuals, which the array combines in the same way as the intrinsic noise~\cite{Hazboun:2019vhv}. Their inter-pulsar correlations enhance this noise by a factor $\approx 3$ for $N_{\rm p}=174$, because the correlated part of the background occupies the same low multipoles, mainly the quadrupole, as the response to a single source.\footnote{This can be made quantitative as follows. The SNR of a source with response vector $\mathbf F$ over the pulsars is $\rho^2 \propto \sum_{IJ} F_I [C^{-1}]_{IJ} F_J$ with $C_{IJ} = S_n\delta_{IJ} + S_h\chi_{IJ}$, where $S_n$ is the intrinsic and $S_h$ the confusion strain-noise spectral density, and $\chi_{IJ} = \tfrac12\delta_{IJ} + \Gamma(\zeta_{IJ})$ is the Hellings--Downs matrix, with $\Gamma(0)=1/2$ so that $\chi_{II}=1$. For $N_{\rm p}$ pulsars spread uniformly over the sky, the spherical harmonics evaluated at the pulsar positions, $Y_{\ell m}(\hat p_I)$, are approximate eigenvectors of $\chi$, with eigenvalues $\lambda_\ell = N_{\rm p} f_\ell/[2(2\ell+1)] + 1/2$, where $f_\ell = 3(2\ell+1)/[(\ell+2)(\ell+1)\ell(\ell-1)]$ are the Hellings--Downs power fractions ($f_2 = 0.625$, $f_3 = 0.175$, \dots, $\sum_\ell f_\ell = 1$). Since $\Gamma$ is the sky average of $F_IF_J$, the source response has the same multipole content, $|\mathbf F_\ell|^2 = f_\ell|\mathbf F|^2$, so that
$$
    \rho^2 \propto |\mathbf F|^2 \sum_{\ell\ge2} \frac{f_\ell}{S_n + \lambda_\ell S_h} \,.
$$
Neglecting the correlations ($\chi_{IJ}\to\delta_{IJ}$), as in Eq.~\eqref{eq:hceff}, gives instead $|\mathbf F|^2/(S_n+S_h)$. In the confusion-dominated regime the correlations therefore enhance the effective confusion noise by $[\sum_\ell f_\ell/\lambda_\ell]^{-1}\approx 2.6$ for $N_{\rm p}=174$. A direct numerical evaluation with the full $\chi$ for random pulsar positions gives $\approx 3$. The enhancement is due to the quadrupole, $\lambda_2 \approx 11$, which the array cannot separate from a single source, whereas the higher multipoles are resolved.} Adding the confusion noise to each pulsar's noise spectrum and combining the array as in Ref.~\cite{Hazboun:2019vhv}, the sensitivity curve becomes
\be \label{eq:hceff}
    h_{\rm c}^{\rm eff}(f)^2 \approx h_{\rm c}^{\rho=1}(f)^2 + \kappa\,\frac{5}{4N_{\rm p}}\, h_{\rm c}^{\rm GW}(f)^2 \,,
\ee
where the factor $5/4$ is the inverse of the inclination and polarization average of the source response that is absorbed in $h_{\rm c}^{\rho=1}$. We adopt $\kappa=3$, while the treatment of Refs.~\cite{Hazboun:2019vhv,SKAOPulsarScienceWorkingGroup:2025oyu}, which neglects the inter-pulsar correlations of the background, corresponds to $\kappa=1$. With Eq.~\eqref{eq:hceff}, the effective threshold amplitude degrades to
\be \label{eq:Amin_eff}
    A_{\rm min}^{\rm eff}(f) = A_{\rm min}(f)
    \sqrt{1 + \kappa\,\frac{5}{4N_{\rm p}}\,\frac{[h_{\rm c}^{\rm GW}(f)]^2}{[h_{\rm c}^{\rho=1}(f)]^2}} \,.
\ee
As Eq.~\eqref{eq:Amin_eff} determines both the resolvability of individual sources and the level of confusion noise, the true value of $A_{\rm min}^{\rm eff}(f)$ is obtained iteratively. The threshold is initially set to $A_{\rm min}^{\rm eff}(f) = A_{\rm min}(f)$, the confusion noise is computed using~\eqref{eq:hcGW}, and the threshold is then updated according to~\eqref{eq:Amin_eff}. The iteration is stopped when the largest fractional change of $A_{\rm min}^{\rm eff}(f)$ across the frequency grid falls below $10^{-3}$, which is typically reached after five iterations. The converged $A_{\rm min}^{\rm eff}(f)$ is independent of the initial guess, indicating that the fixed point is unique. Since the effect of the confusion noise tends to be modest, we find that a single iteration estimates $A_{\rm min}^{\rm eff}(f)$ with $\mathcal{O}(10\%)$ accuracy and two are sufficient for a $\mathcal{O}(1\%)$ accuracy.

\begin{figure}
    \centering
    \includegraphics[width=\linewidth]{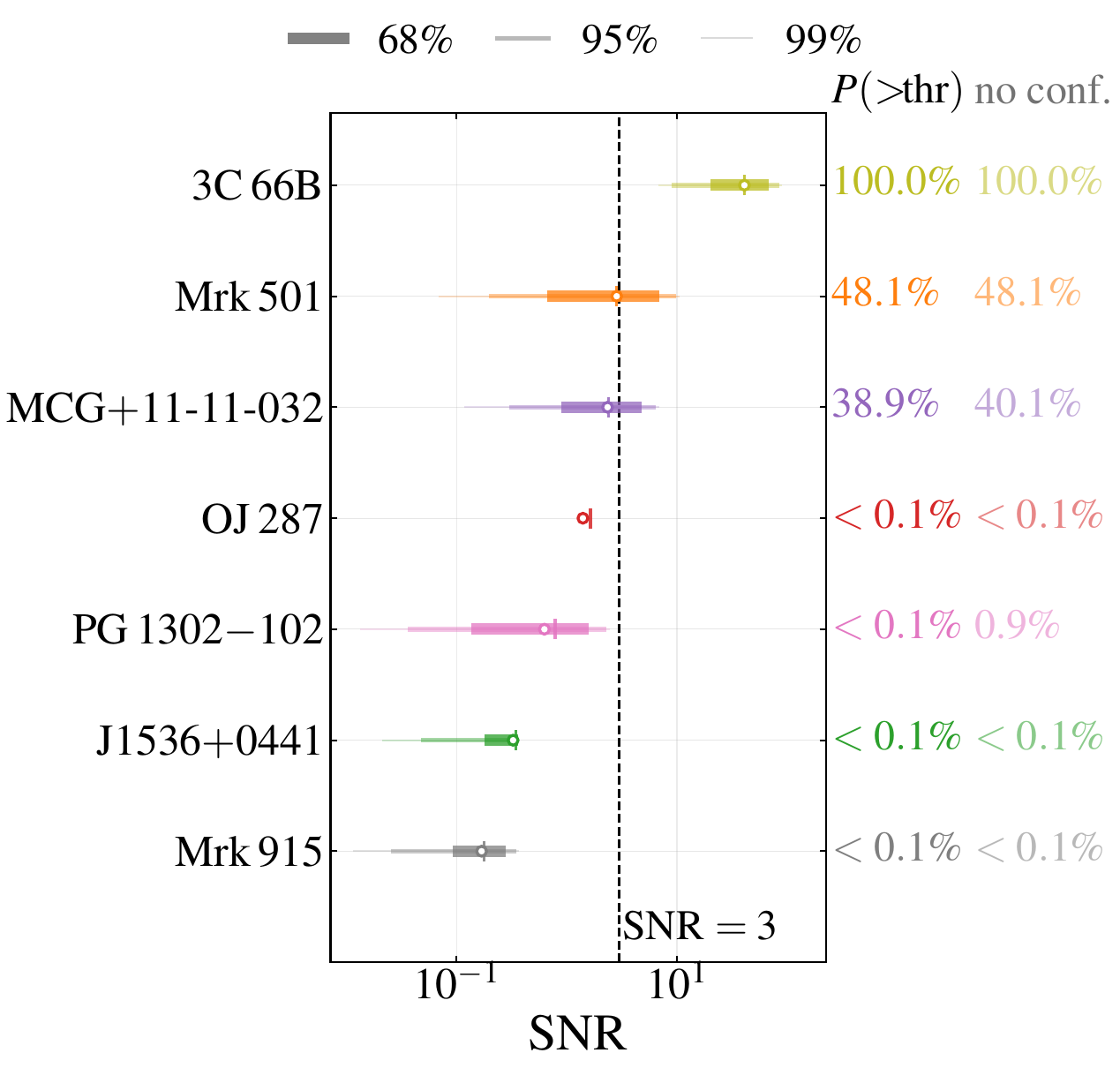}
    \caption{SNRs $\rho = A^{(1)}/A_{\rm min}^{\rm eff}$ of the EM-identified candidates of Table~\ref{tab:candidates} for SKAO, including the confusion noise from unresolved binaries. The bars show the $68\%$, $95\%$ and $99\%$ credible ranges obtained by mapping the PTA posteriors of Fig.~\ref{fig:corner} onto $A_{\rm min}^{\rm eff}$ and marginalizing over the candidate mass uncertainties as in Fig.~\ref{fig:expected}, with the open circle marking the median. The vertical tick gives the median obtained when the confusion noise is neglected. The dashed line marks the detection threshold $\rho_{\rm th}=3$, and the right-hand axis quotes the probability that the candidate exceeds it.}
    \label{fig:snrs}
\end{figure}

\begin{figure}
    \centering
    \includegraphics[width=0.96\linewidth]{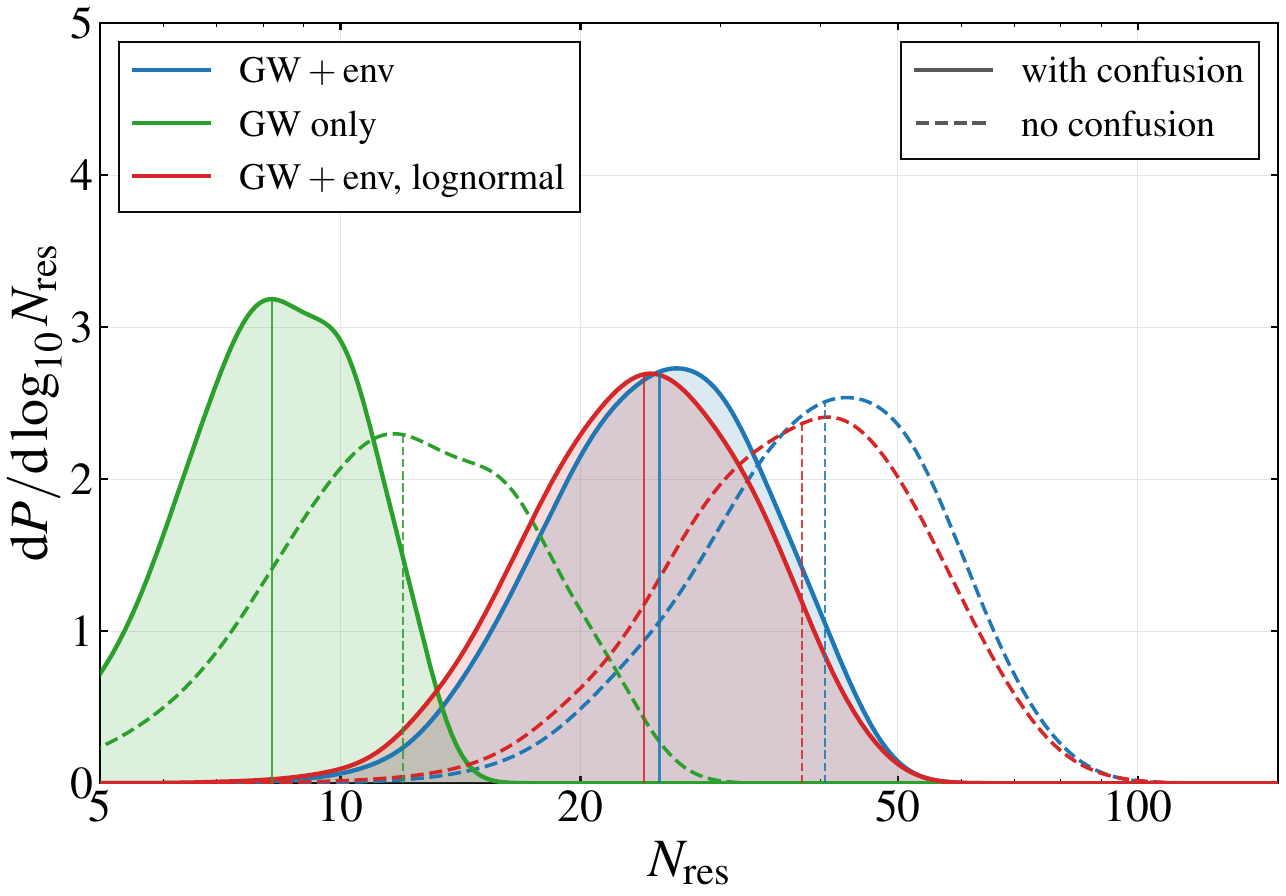}
    \caption{Posterior distribution of the expected number of individually resolvable sources $\bar N_{\rm res}$, Eq.~\eqref{eq:Nres}, for SKAO, obtained by mapping the NANOGrav 15-year posteriors of Fig.~\ref{fig:corner} onto $\bar N_{\rm res}$. The solid curves include the confusion noise from the unresolved binaries while the dashed curves use only the bare SKAO sensitivity. Blue and green show the GW+environment and GW-only models obtained with the variance-averaged likelihood, and red the GW+environment model obtained with the log-normal approximation. The vertical lines mark the medians.}
    \label{fig:Nres}
\end{figure}

Fig.~\ref{fig:snrs} shows the resulting SNRs of the candidate sources, accounting for the confusion noise from the unresolved binaries. 3C~66B is the only robust target, as its entire range lies above the threshold, with a median $\rho \simeq 40$. It would be resolved by SKAO for every combination of candidate masses and population parameters allowed by the NANOGrav 15-year fit. Mrk~501 and MCG+11-11-032 straddle the threshold, with $P(\rho>\rho_{\rm th}) = 48.1\%$ and $38.9\%$, their broad distributions reflecting the poorly constrained total masses of both systems. The remaining candidates fall well below the threshold, with medians $\rho \lesssim 1$. A comparison with the medians obtained when the confusion noise is neglected shows that the unresolved population degrades the candidate SNRs only mildly, and that the degradation grows towards lower frequencies. In particular, it is largest for PG~1302-102 at $12\,$nHz, and negligible for 3C~66B at $60\,$nHz. Nevertheless, the effect acts in the direction of pushing marginal sources, such as Mrk~501, below the threshold, and it is the reason why the resolvable population of Fig.~\ref{fig:Nres} is substantially smaller than the naive estimate.

MCG+11-11-032 also illustrates the difference between rarity and resolvability. It reaches the threshold in $38.9\%$ of the posterior samples, yet $P(0)=0.05$ implies that many louder binaries are expected above its frequency, placing it in a crowded part of the population. However, when considering only the frequency bin around its GW frequency 29~nHz find that $P(0)=0.48$ (or $P(0)=0.79$) in the GW+env (GW-only) model corresponding to less than a single expected event in both models, so it is expected that the event can be identified by SKAO.

\begin{figure*}
    \centering
    \includegraphics[width=0.96\linewidth]{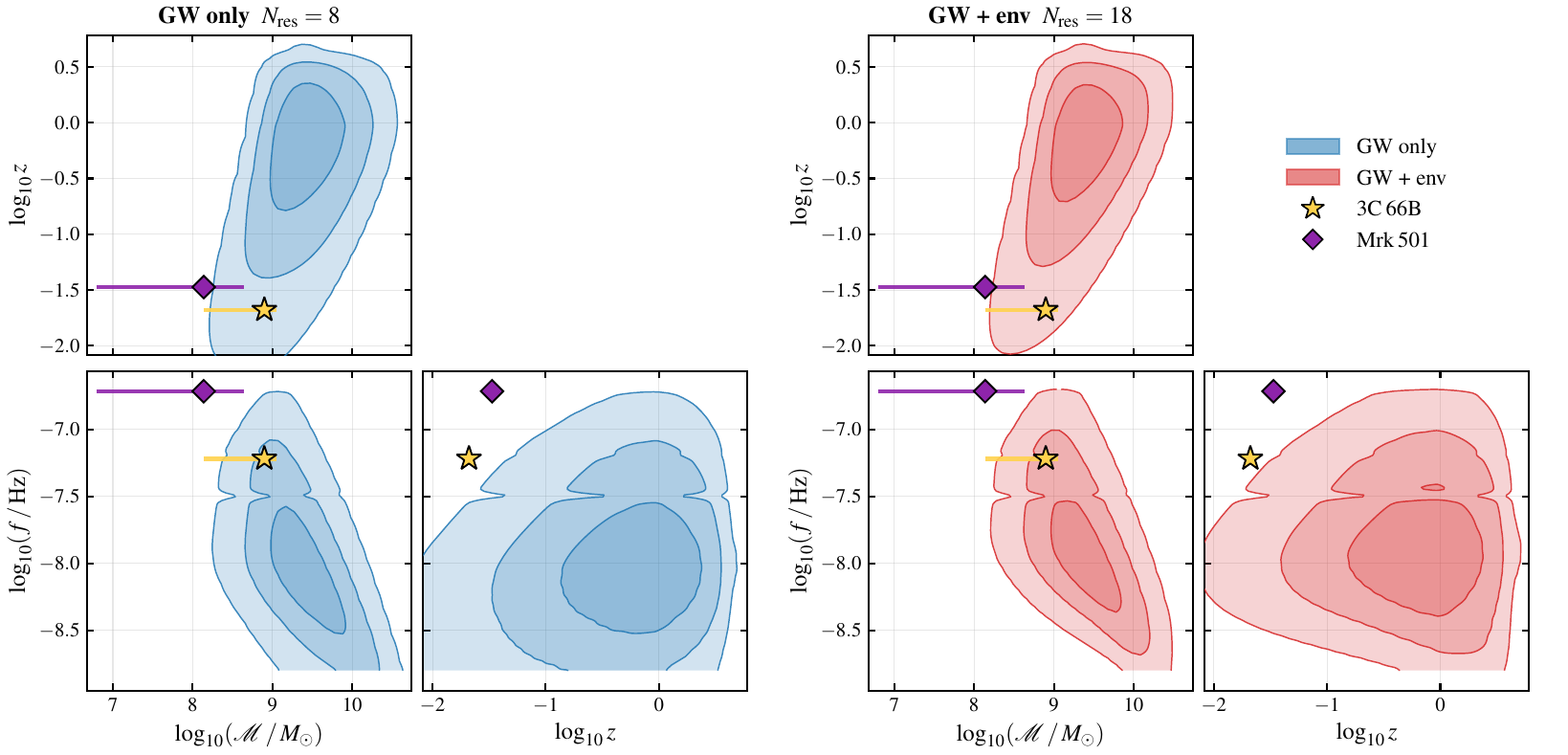}
    \caption{Forecast distribution of SMBH binaries individually resolvable by SKAO in redshift $z$, chirp mass $\mathcal{M}$, and observed GW frequency $f$ for the best fit parameters of the PTA analysis. The left panels show the GW-only model and the right panels the GW+environment model, with the total expected number of resolvable sources $\bar{N}_{\rm res}$ indicated in each title and the colour scale giving the number of resolvable events per pixel. The symbols show the strongest candidate sources 3C~66B and Mrk~501.}
    \label{fig:ska_events}
\end{figure*}

To quantify the population detectable with SKAO, we compute the expected number of resolvable sources by integrating the GWAD over $A > \rho_{\rm th}A_{\rm min}^{\rm eff}(f)$:
\be \label{eq:Nres}
    \bar{N}_{\rm res} = \int \td \ln f \int_{\rho_{\rm th} A_{\rm min}^{\rm eff}(f)} \td A \, \frac{\td N}{\td A \,\td \ln f} \,.
\ee
Figs.~\ref{fig:corner} and~\ref{fig:Nres} show the distribution of $\bar{N}_{\rm res}$ obtained by mapping the PTA posterior onto the expected number of resolvable sources. We find $\bar{N}_{\rm res} = 8^{+2}_{-2}$ in the purely GW-driven model, compared with the more than three times larger $\bar{N}_{\rm res} = 25^{+9}_{-7}$ in the model including environmental hardening, consistent with the higher formation probability and enhanced high-frequency abundance discussed above. These values are also broadly consistent with the number of expected resolved sources found by the SKAO collaboration~\cite{SKAOPulsarScienceWorkingGroup:2025oyu} and other earlier studies~\cite{Sesana:2008xk, Rosado:2015epa, Mingarelli:2017fbe, Kelley:2017vox, Becsy:2022pnr}. In Fig.~\ref{fig:corner}, the number of resolvable sources $\bar{N}_{\rm res}$ is seen to correlate strongly with the overall normalization of the merger rate $p_{\rm BH}$, and it shows a milder but negative correlation with the time of $t_{\rm ref}$. This is expected, because $t_{\rm ref}$ controls the timescale of environmental hardening and $t_{\rm ref}\to \infty$ recovers the GW-driven case for which the expectation for $\bar{N}_{\rm res}$ is lower.

We also find that neglecting the confusion noise from unresolved binaries leads to an overestimate of $\bar{N}_{\rm res}$, by roughly 60\%, $\bar{N}_{\rm res} = 25^{+9}_{-7}$ and $40^{+15}_{-13}$ in the two noise treatments respectively, as shown by the dashed distributions. 

To quantify the properties of resolvable binaries, Fig.~\ref{fig:ska_events} shows their distribution in redshift, chirp mass $\mathcal{M}$, and observed frequency $f$ for the best fit parameters of the PTA analysis in the GW-driven model (left panel) and the model including environmental effects (right panel). The resolvable events cluster at high mass ($\mathcal{M} \sim 10^{9}-10^{10}\,\Msun$) and low redshift ($z \lesssim 1$), reflecting the steep amplitude scaling of Eq.~\eqref{eq:A1}, at frequencies of a few to $\sim 10\,$nHz. The candidates 3C~66B and Mrk~501 reside at the high-frequency and low-$z$ corner of the detectable population.

\section{Conclusions}
\label{sec:concl}

We have assessed whether electromagnetically identified SMBH binary candidates occupy the rare high-amplitude tail of the GW amplitude distribution, using a population model fitted to the NANOGrav 15-year data. The GW amplitude distribution provides a single framework in which the bulk of the distribution sets the SGWB and its tail sets the individually resolvable sources, so that the same fit describes both. We estimated the expected number $\bar N$ of binaries that are louder and at higher frequency for the three most promising candidates. 3C~66B and Mrk~501 are potential rare outliers, with probabilities $P(0)=0.96$ and $0.81$ of being the loudest source above their frequency in the environmentally driven model, and $0.99$ and $0.95$ for purely GW-driven evolution, whereas MCG+11-11-032, with $P(0)=0.05$, is a typical member of the population. 

The comparison between the candidates and the population carries information on binary evolution that the SGWB spectrum alone does not. Environmental hardening drives binaries faster through the low-frequency regime, so the fit compensates with a binary formation efficiency about six times larger than in the purely GW-driven case, which in turn populates the high-frequency loud tail. Both scenarios are compatible with the NANOGrav spectrum, yet they differ by more than a factor of three in the number of binaries SKAO would resolve, $\bar N_{\rm res}\simeq 25$ against $8$, and by a factor of six in the expected number of sources louder than 3C~66B. The abundance of resolved individual sources is therefore a probe of the hardening mechanism complementary to the spectral shape, and a confirmed loud source such as 3C~66B would favor environmentally driven evolution. Realizing this requires firmer EM characterization, since the candidate periods and masses remain uncertain, non-binary origins of the observed periodicities are not excluded, and eccentricity, which we have omitted in this analysis, would populate the high-frequency tail further.

For SKAO, we forecast tens of individually resolvable binaries, concentrated at chirp masses of $10^{9}-10^{10}\,\Msun$, redshifts $z\lesssim1$, and frequencies of a few to $\sim10\,$nHz. The confusion noise from the unresolved population lowers this number by about $40\%$. Among the current candidates, 3C~66B is the only one robustly above the SKAO threshold, with a median signal-to-noise ratio of $\simeq40$ and no posterior samples below threshold, while Mrk~501 and MCG+11-11-032 straddle it. Current PTA limits already reach into the upper part of its allowed mass range~\cite{Tremblay:2025fgk}, and since no posterior sample falls below the SKAO threshold, SKAO will either detect 3C~66B or exclude the binary interpretation across the mass range considered here. More generally, the resolvable population predicted here marks the region where EM searches for counterparts should concentrate, namely nearby galaxies hosting black holes of $\gtrsim10^{9}\,\Msun$ with orbital periods of roughly $5$--$20$\,yr. As SKAO resolves individual binaries and the candidate sample grows, the joint analysis of individual sources and the SGWB within the GWAD framework will strengthen the inference on the binary formation efficiency and the hardening mechanism, and identified counterparts would open the way to standard-siren measurements of the local expansion history.

\vspace{10pt}
\noindent
\emph{Acknowledgments:} This work was supported by the Estonian Research Council grants PSG869, TARISTU24-TK3, TARISTU24-TK10, and the Centre of Excellence programme TK202 of the Estonian Ministry of Education and Research. 

\bibliography{refs}

@article{Agarwal:2026dlv,
    author = "Agarwal, Nikita and others",
    title = "{The NANOGrav 15 yr Data Set: Impacts of Customized Chromatic Noise Models on Gravitational Wave Analyses}",
    eprint = "2606.28554",
    archivePrefix = "arXiv",
    primaryClass = "astro-ph.CO",
    month = "6",
    year = "2026",
    journal = "arXiv preprint"
}

@article{Larsen:2026pru,
    author = "Larsen, Bjorn and others",
    title = "{The NANOGrav 15 yr Data Set: Customized Chromatic Noise Models}",
    eprint = "2606.28571",
    archivePrefix = "arXiv",
    primaryClass = "astro-ph.HE",
    doi = "10.3847/1538-4357/ae677c",
    journal = "Astrophys. J.",
    volume = "1005",
    pages = "29",
    year = "2026"
}

@article{Keitaanranta:2025ncl,
    author = "Keitaanranta, Atte and Johansson, Peter H. and Rawlings, Alexander and Tuominen, Toni and Rantala, Antti and Naab, Thorsten and Liao, Shihong and Reinoso, Basti{\'a}n",
    title = "{Rapid sinking and efficient mergers of supermassive black holes in compact high-redshift galaxies}",
    eprint = "2512.11665",
    archivePrefix = "arXiv",
    primaryClass = "astro-ph.GA",
    doi = "10.1093/mnras/stag756",
    journal = "Mon. Not. Roy. Astron. Soc.",
    volume = "549",
    number = "1",
    pages = "stag756",
    year = "2026"
}

@article{Hazboun:2019vhv,
    author = "Hazboun, Jeffrey S. and Romano, Joseph D. and Smith, Tristan L.",
    title = "{Realistic sensitivity curves for pulsar timing arrays}",
    eprint = "1907.04341",
    archivePrefix = "arXiv",
    primaryClass = "gr-qc",
    doi = "10.1103/PhysRevD.100.104028",
    journal = "Phys. Rev. D",
    volume = "100",
    number = "10",
    pages = "104028",
    year = "2019"
}

@article{Mingarelli:2017fbe,
    author = "Mingarelli, Chiara M. F. and Lazio, T. Joseph W. and Sesana, Alberto and Greene, Jenny E. and Ellis, Justin A. and Ma, Chung-Pei and Croft, Steve and Burke-Spolaor, Sarah and Taylor, Stephen R.",
    title = "{The Local Nanohertz Gravitational-Wave Landscape From Supermassive Black Hole Binaries}",
    eprint = "1708.03491",
    archivePrefix = "arXiv",
    primaryClass = "astro-ph.GA",
    doi = "10.1038/s41550-017-0299-6",
    journal = "Nature Astron.",
    volume = "1",
    number = "12",
    pages = "886--892",
    year = "2017"
}

@article{Quinlan:1996vp,
    author = "Quinlan, Gerald D.",
    title = "{The dynamical evolution of massive black hole binaries - I. hardening in a fixed stellar background}",
    eprint = "astro-ph/9601092",
    archivePrefix = "arXiv",
    reportNumber = "RUTGERS-ASTROPHYSICS-PREPRINT-SERIES-NO-187",
    doi = "10.1016/S1384-1076(96)00003-6",
    journal = "New Astron.",
    volume = "1",
    pages = "35--56",
    year = "1996"
}

@article{Shakura:1972te,
    author = "Shakura, N. I. and Sunyaev, R. A.",
    title = "{Black holes in binary systems. Observational appearance}",
    journal = "Astron. Astrophys.",
    volume = "24",
    pages = "337--355",
    year = "1973"
}

@article{Kormendy:2013dxa,
    author = "Kormendy, John and Ho, Luis C.",
    title = "{Coevolution (Or Not) of Supermassive Black Holes and Host Galaxies}",
    eprint = "1304.7762",
    archivePrefix = "arXiv",
    primaryClass = "astro-ph.CO",
    doi = "10.1146/annurev-astro-082708-101811",
    journal = "Ann. Rev. Astron. Astrophys.",
    volume = "51",
    pages = "511--653",
    year = "2013"
}

@article{Kocsis:2012ui,
    author = "Kocsis, Bence and Haiman, Zoltan and Loeb, Abraham",
    title = "{Gas pile-up, gap overflow, and Type 1.5 migration in circumbinary disks: application to supermassive black hole binaries}",
    eprint = "1205.5268",
    archivePrefix = "arXiv",
    primaryClass = "astro-ph.HE",
    doi = "10.1111/j.1365-2966.2012.22118.x",
    journal = "Mon. Not. Roy. Astron. Soc.",
    volume = "427",
    pages = "2680--2700",
    year = "2012"
}

@article{Kocsis:2010xa,
    author = "Kocsis, Bence and Sesana, Alberto",
    title = "{Gas driven massive black hole binaries: signatures in the nHz gravitational wave background}",
    eprint = "1002.0584",
    archivePrefix = "arXiv",
    primaryClass = "astro-ph.CO",
    doi = "10.1111/j.1365-2966.2010.17782.x",
    journal = "Mon. Not. Roy. Astron. Soc.",
    volume = "411",
    pages = "1467",
    year = "2011"
}

@article{Syer:1995hk,
    author = "Syer, Dave and Clarke, Catherine J.",
    title = "{Satellites in discs: regulating the accretion luminosity}",
    eprint = "astro-ph/9505021",
    archivePrefix = "arXiv",
    reportNumber = "SYER95-2",
    doi = "10.1093/mnras/277.3.758",
    journal = "Mon. Not. Roy. Astron. Soc.",
    volume = "277",
    pages = "758",
    year = "1995"
}

@article{Haiman:2009te,
    author = "Haiman, Zolt{\'a}n and Haiman, Zoltan and Kocsis, Bence and Kocsis, Bence and Menou, Kristen and Menou, Kristen",
    title = "{The Population of Viscosity- and Gravitational Wave-Driven Supermassive Black Hole Binaries Among Luminous AGN}",
    eprint = "0904.1383",
    archivePrefix = "arXiv",
    primaryClass = "astro-ph.CO",
    doi = "10.1088/0004-637X/700/2/1952",
    journal = "Astrophys. J.",
    volume = "700",
    pages = "1952--1969",
    year = "2009",
    note = "[Erratum: Astrophys.J. 937, 129 (2022)]"
}

@article{Chandrasekhar:1943ys,
    author = "Chandrasekhar, Subrahmanyan",
    title = "{Dynamical Friction. I. General Considerations: the Coefficient of Dynamical Friction}",
    doi = "10.1086/144517",
    journal = "Astrophys. J.",
    volume = "97",
    pages = "255",
    year = "1943"
}

@article{Ivanov:1998qk,
    author = "Ivanov, P. B. and Papaloizou, J. C. B. and Polnarev, A. G.",
    title = "{The evolution of a supermassive binary caused by an accretion disc}",
    eprint = "astro-ph/9812198",
    archivePrefix = "arXiv",
    reportNumber = "TAC-1998-027",
    doi = "10.1046/j.1365-8711.1999.02623.x",
    journal = "Mon. Not. Roy. Astron. Soc.",
    volume = "307",
    pages = "79",
    year = "1999"
}

@article{Becsy:2022pnr,
    author = "B{\'e}csy, Bence and Cornish, Neil J. and Kelley, Luke Zoltan",
    title = "{Exploring Realistic Nanohertz Gravitational-wave Backgrounds}",
    eprint = "2207.01607",
    archivePrefix = "arXiv",
    primaryClass = "astro-ph.HE",
    doi = "10.3847/1538-4357/aca1b2",
    journal = "Astrophys. J.",
    volume = "941",
    number = "2",
    pages = "119",
    year = "2022"
}

@article{Jenet:2003ew,
    author = "Jenet, Fredrick A. and Lommen, Andrea and Larson, Shane L. and Wen, Linqing",
    title = "{Constraining the properties of the proposed supermassive black hole system in 3c66b: Limits from pulsar timing}",
    eprint = "astro-ph/0310276",
    archivePrefix = "arXiv",
    doi = "10.1086/383020",
    journal = "Astrophys. J.",
    volume = "606",
    pages = "799--803",
    year = "2004"
}

@article{NANOGrav:2023wsz,
    author = "Agazie, Gabriella and others",
    collaboration = "NANOGrav",
    title = "{The NANOGrav 12.5 yr Data Set: A Computationally Efficient Eccentric Binary Search Pipeline and Constraints on an Eccentric Supermassive Binary Candidate in 3C 66B}",
    eprint = "2309.17438",
    archivePrefix = "arXiv",
    primaryClass = "astro-ph.HE",
    doi = "10.3847/1538-4357/ad1f61",
    journal = "Astrophys. J.",
    volume = "963",
    number = "2",
    pages = "144",
    year = "2024"
}

@article{SKAOPulsarScienceWorkingGroup:2025oyu,
    author = "Shannon, Ryan M. and others",
    collaboration = "SKAO Pulsar Science Working Group",
    title = "{The SKAO Pulsar Timing Array}",
    eprint = "2512.16163",
    archivePrefix = "arXiv",
    primaryClass = "astro-ph.HE",
    journal = "arXiv preprint",
    month = "12",
    year = "2025"
}

@article{Janssen:2014dka,
    author = "Janssen, Gemma and others",
    editor = "Bourke, Tyler L. and others",
    title = "{Gravitational wave astronomy with the SKA}",
    eprint = "1501.00127",
    archivePrefix = "arXiv",
    primaryClass = "astro-ph.IM",
    doi = "10.22323/1.215.0037",
    journal = "PoS",
    volume = "AASKA14",
    pages = "037",
    year = "2015"
}

@article{Sardana:2026jvr,
    author = "Sardana, Shubhit and Goncharov, Boris and Cardinal Tremblay, Jacob",
    title = "{The Targeted Standard Siren Cosmology with Pulsar Timing Arrays}",
    eprint = "2603.12168",
    archivePrefix = "arXiv",
    primaryClass = "astro-ph.CO",
    journal = "arXiv preprint",
    month = "3",
    year = "2026"
}

@article{Wang:2022oou,
    author = "Wang, Ling-Feng and Shao, Yue and Xiao, Si-Ren and Zhang, Jing-Fei and Zhang, Xin",
    title = "{Ultra-low-frequency gravitational waves from individual supermassive black hole binaries as standard sirens}",
    eprint = "2201.00607",
    archivePrefix = "arXiv",
    primaryClass = "astro-ph.CO",
    doi = "10.1088/1475-7516/2025/05/095",
    journal = "JCAP",
    volume = "05",
    pages = "095",
    year = "2025"
}

@article{Planck:2018vyg,
    author = "Aghanim, N. and others",
    collaboration = "Planck",
    title = "{Planck 2018 results. VI. Cosmological parameters}",
    eprint = "1807.06209",
    archivePrefix = "arXiv",
    primaryClass = "astro-ph.CO",
    doi = "10.1051/0004-6361/201833910",
    journal = "Astron. Astrophys.",
    volume = "641",
    pages = "A6",
    year = "2020",
    note = "[Erratum: Astron.Astrophys. 652, C4 (2021)]"
}

@article{Tremblay:2025fgk,
    author = "Tremblay, Jacob Cardinal and others",
    title = "{A Multi-Messenger Search for the Supermassive Black Hole Binary in 3C 66B with the Parkes Pulsar Timing Array}",
    eprint = "2508.20007",
    archivePrefix = "arXiv",
    primaryClass = "astro-ph.HE",
    doi = "10.3847/2041-8213/ae3c98",
    journal = "Astrophys. J. Lett.",
    volume = "998",
    number = "2",
    pages = "L42",
    year = "2026"
}

@article{Britzen:2026stag,
    author = {Britzen, S and Olivares, H and Gopal-Krishna and Jaron, F and Pashchenko, I N and Kun, E and Schinzel, F K and González, J Becerra and Paneque, D and MacDonald, N R},
    title = {{Detection of a second jet within the nuclear core of Mrk 501}},
    doi = "10.1093/mnras/stag291",
    journal = "Mon. Not. Roy. Astron. Soc.",
    volume = "548",
    number = "3",
    pages = "1568--1582",
    year = "2026"
}

@article{Raidal:2026ezm,
    author = {Raidal, Juhan and Urrutia, Juan and Vaskonen, Ville and Veerm{\"a}e, Hardi},
    title = "{The Heavy Tailed Non-Gaussianity of the Supermassive Black Hole Gravitational Wave Background}",
    eprint = "2604.08506",
    archivePrefix = "arXiv",
    primaryClass = "astro-ph.CO",
    month = "4",
    journal = "arXiv preprint",
    year = "2026"
}

@article{DeRosa:2019myq,
    author = "De Rosa, Alessandra and others",
    title = "{The quest for dual and binary supermassive black holes: A multi-messenger view}",
    eprint = "2001.06293",
    archivePrefix = "arXiv",
    primaryClass = "astro-ph.GA",
    doi = "10.1016/j.newar.2020.101525",
    journal = "New Astron. Rev.",
    volume = "86",
    pages = "101525",
    year = "2019"
}

@article{Severgnini:2018jol,
    author = "Severgnini, P. and others",
    title = "{Swift data hint at a binary supermassive black hole candidate at sub-parsec separation}",
    eprint = "1806.10150",
    archivePrefix = "arXiv",
    primaryClass = "astro-ph.HE",
    doi = "10.1093/mnras/sty1699",
    journal = "Mon. Not. Roy. Astron. Soc.",
    volume = "479",
    number = "3",
    pages = "3804--3813",
    year = "2018"
}

@article{NANOGrav:2025gqp,
    author = "Agarwal, Nikita and others",
    collaboration = "NANOGrav",
    title = "{The NANOGrav 15 yr Dataset: Targeted Searches for Supermassive Black Hole Binaries}",
    eprint = "2508.16534",
    archivePrefix = "arXiv",
    primaryClass = "astro-ph.HE",
    doi = "10.3847/2041-8213/ae3719",
    journal = "Astrophys. J. Lett.",
    volume = "998",
    number = "1",
    pages = "L11",
    year = "2026"
}

@ARTICLE{1996ApJ...460..207L,
       author = {{Lehto}, Harry J. and {Valtonen}, Mauri J.},
        title = "{OJ 287 Outburst Structure and a Binary Black Hole Model}",
      journal = {\apj},
         year = 1996,
        month = mar,
       volume = {460},
        pages = {207},
          doi = {10.1086/176962},
       adsurl = {https://ui.adsabs.harvard.edu/abs/1996ApJ...460..207L}
}

@article{Laine:2020dnr,
    author = "Laine, Seppo and others",
    title = "{Spitzer Observations of the Predicted Eddington Flare from Blazar OJ 287}",
    eprint = "2004.13392",
    archivePrefix = "arXiv",
    primaryClass = "astro-ph.HE",
    doi = "10.3847/2041-8213/ab79a4",
    journal = "Astrophys. J. Lett.",
    volume = "894",
    number = "1",
    pages = "L1",
    year = "2020"
}

@article{Sudou:2003hv,
    author = "Sudou, Hiroshi and Iguchi, Satoru and Murata, Yasuhiro and Taniguchi, Yoshiaki",
    title = "{Orbital motion in the radio galaxy 3c 66b: evidence for a supermassive black hole binary}",
    eprint = "astro-ph/0306103",
    archivePrefix = "arXiv",
    doi = "10.1126/science.1082817",
    journal = "Science",
    volume = "300",
    pages = "1263",
    year = "2003"
}

@article{Ghisellini:2009fj,
    author = "Ghisellini, G. and Tavecchio, F. and Foschini, L. and Ghirlanda, G. and Maraschi, L. and Celotti, A.",
    title = "{General physical properties of bright Fermi blazars}",
    eprint = "0909.0932",
    archivePrefix = "arXiv",
    primaryClass = "astro-ph.CO",
    doi = "10.1111/j.1365-2966.2009.15898.x",
    journal = "Mon. Not. Roy. Astron. Soc.",
    volume = "402",
    pages = "497",
    year = "2010"
}

@ARTICLE{2010ApJ...724L.166I,
       author = {{Iguchi}, Satoru and {Okuda}, Takeshi and {Sudou}, Hiroshi},
        title = "{A Very Close Binary Black Hole in a Giant Elliptical Galaxy 3C 66B and its Black Hole Merger}",
      journal = {\apjl},
         year = 2010,
        month = dec,
       volume = {724},
       number = {2},
        pages = {L166-L170},
          doi = {10.1088/2041-8205/724/2/L166},
archivePrefix = {arXiv},
       eprint = {1011.2647},
 primaryClass = {astro-ph.GA},
       adsurl = {https://ui.adsabs.harvard.edu/abs/2010ApJ...724L.166I}
}

@article{NANOGrav:2020lwu,
    author = "Arzoumanian, Zaven and others",
    collaboration = "NANOGrav",
    title = "{Multimessenger Gravitational-wave Searches with Pulsar Timing Arrays: Application to 3C 66B Using the NANOGrav 11-year Data Set}",
    eprint = "2005.07123",
    archivePrefix = "arXiv",
    primaryClass = "astro-ph.GA",
    doi = "10.3847/1538-4357/ababa1",
    journal = "Astrophys. J.",
    volume = "900",
    number = "2",
    pages = "102",
    year = "2020"
}

@article{Graham:2015gma,
    author = "Graham, Matthew J. and Djorgovski, S. George and Stern, Daniel and Glikman, Eilat and Drake, Andrew J. and Mahabal, Ashish A. and Donalek, Ciro and Larson, Steve and Christensen, Eric",
    title = "{A possible close supermassive black-hole binary in a quasar with optical periodicity}",
    eprint = "1501.01375",
    archivePrefix = "arXiv",
    primaryClass = "astro-ph.GA",
    doi = "10.1038/nature14143",
    journal = "Nature",
    volume = "518",
    pages = "74",
    year = "2015"
}

@article{Kozhikkal:2023gkt,
    author = "Kozhikkal, Musfar Muhamed and Chen, Siyuan and Theureau, Gilles and Habouzit, Melanie and Sesana, Alberto",
    title = "{Mass-redshift dependency of supermassive black hole binaries for the gravitational wave background}",
    eprint = "2305.18293",
    archivePrefix = "arXiv",
    primaryClass = "astro-ph.CO",
    doi = "10.1093/mnras/stae1219",
    journal = "Mon. Not. Roy. Astron. Soc.",
    volume = "531",
    number = "1",
    pages = "1931--1950",
    year = "2024"
}

@ARTICLE{2009ApJ...700.1952H,
       author = {{Haiman}, Zolt{\'a}n and {Kocsis}, Bence and {Menou}, Kristen},
        title = "{The Population of Viscosity- and Gravitational Wave-driven Supermassive Black Hole Binaries Among Luminous Active Galactic Nuclei}",
      journal = {\apj},
         year = 2009,
        month = aug,
       volume = {700},
       number = {2},
        pages = {1952-1969},
          doi = {10.1088/0004-637X/700/2/1952},
archivePrefix = {arXiv},
       eprint = {0904.1383},
 primaryClass = {astro-ph.CO},
       adsurl = {https://ui.adsabs.harvard.edu/abs/2009ApJ...700.1952H}
}

@article{Serafinelli:2020nog,
    author = "Serafinelli, Roberto and others",
    title = "{Unveiling Sub-pc Supermassive Black Hole Binary Candidates in Active Galactic Nuclei}",
    eprint = "2009.06662",
    archivePrefix = "arXiv",
    primaryClass = "astro-ph.HE",
    doi = "10.3847/1538-4357/abb3c3",
    journal = "Astrophys. J.",
    volume = "902",
    number = "1",
    pages = "10",
    year = "2020"
}

@misc{Raidal_GWAD_2026,
    author = {Raidal, Juhan and Urrutia, Juan and Vaskonen, Ville and Veermäe, Hardi},
    license = {GPL-3.0-or-later},
    title = {{GWADpy}},
    version = {1.0.0},
    url = {https://github.com/JRaidal/GWADpy},
    year = {2026}
}

@article{NANOGrav:2023hfp,
    author = "Agazie, Gabriella and others",
    collaboration = "NANOGrav",
    title = "{The NANOGrav 15 yr Data Set: Constraints on Supermassive Black Hole Binaries from the Gravitational-wave Background}",
    eprint = "2306.16220",
    archivePrefix = "arXiv",
    primaryClass = "astro-ph.HE",
    doi = "10.3847/2041-8213/ace18b",
    journal = "Astrophys. J. Lett.",
    volume = "952",
    number = "2",
    pages = "L37",
    year = "2023"
}

@ARTICLE{2013CQGra..30x4005M,
       author = {{Merritt}, David},
        title = "{Loss-cone dynamics}",
      journal = {Classical and Quantum Gravity},
         year = 2013,
        month = dec,
       volume = {30},
       number = {24},
          eid = {244005},
        pages = {244005},
          doi = {10.1088/0264-9381/30/24/244005},
archivePrefix = {arXiv},
       eprint = {1307.3268},
 primaryClass = {astro-ph.GA},
       adsurl = {https://ui.adsabs.harvard.edu/abs/2013CQGra..30x4005M}
}

@article{Kelley:2016gse,
    author = "Kelley, Luke Zoltan and Blecha, Laura and Hernquist, Lars",
    title = "{Massive Black Hole Binary Mergers in Dynamical Galactic Environments}",
    eprint = "1606.01900",
    archivePrefix = "arXiv",
    primaryClass = "astro-ph.HE",
    doi = "10.1093/mnras/stw2452",
    journal = "Mon. Not. Roy. Astron. Soc.",
    volume = "464",
    number = "3",
    pages = "3131--3157",
    year = "2017"
}

@article{Mingarelli:2026kjw,
    author = "Mingarelli, Chiara M. F. and Larsen, Bjorn and Eisenberg, Ellis and Zheng, Qinyuan and Hutchison, Forrest",
    title = "{Fingerprints of Individual Supermassive Black Hole Binaries in Pulsar Timing Arrays}",
    eprint = "2603.05722",
    archivePrefix = "arXiv",
    primaryClass = "astro-ph.HE",
    month = "3",
    journal = "arXiv preprint",
    year = "2026"
}

@article{Sesana:2008xk,
    author = "Sesana, A. and Vecchio, A. and Volonteri, M.",
    title = "{Gravitational waves from resolvable massive black hole binary systems and observations with Pulsar Timing Arrays}",
    eprint = "0809.3412",
    archivePrefix = "arXiv",
    primaryClass = "astro-ph",
    doi = "10.1111/j.1365-2966.2009.14499.x",
    journal = "Mon. Not. Roy. Astron. Soc.",
    volume = "394",
    pages = "2255",
    year = "2009"
}

@article{Rosado:2015epa,
    author = "Rosado, Pablo A. and Sesana, Alberto and Gair, Jonathan",
    title = "{Expected properties of the first gravitational wave signal detected with pulsar timing arrays}",
    eprint = "1503.04803",
    archivePrefix = "arXiv",
    primaryClass = "astro-ph.HE",
    doi = "10.1093/mnras/stv1098",
    journal = "Mon. Not. Roy. Astron. Soc.",
    volume = "451",
    number = "3",
    pages = "2417--2433",
    year = "2015"
}

@article{Wyithe:2002ep,
    author = "Wyithe, J. Stuart B. and Loeb, Abraham",
    title = "{Low - frequency gravitational waves from massive black hole binaries: Predictions for LISA and pulsar timing arrays}",
    eprint = "astro-ph/0211556",
    archivePrefix = "arXiv",
    doi = "10.1086/375187",
    journal = "Astrophys. J.",
    volume = "590",
    pages = "691--706",
    year = "2003"
}

@article{Rajagopal:1994zj,
    author = "Rajagopal, Mohan and Romani, Roger W.",
    title = "{Ultralow frequency gravitational radiation from massive black hole binaries}",
    eprint = "astro-ph/9412038",
    archivePrefix = "arXiv",
    doi = "10.1086/175813",
    journal = "Astrophys. J.",
    volume = "446",
    pages = "543--549",
    year = "1995"
}

@article{Sesana:2004sp,
    author = "Sesana, Alberto and Haardt, Francesco and Madau, Piero and Volonteri, Marta",
    title = "{Low - frequency gravitational radiation from coalescing massive black hole binaries in hierarchical cosmologies}",
    eprint = "astro-ph/0401543",
    archivePrefix = "arXiv",
    doi = "10.1086/422185",
    journal = "Astrophys. J.",
    volume = "611",
    pages = "623--632",
    year = "2004"
}

@article{Kelley:2017vox,
    author = "Kelley, Luke Zoltan and Blecha, Laura and Hernquist, Lars and Sesana, Alberto and Taylor, Stephen R.",
    title = "{Single Sources in the Low-Frequency Gravitational Wave Sky: properties and time to detection by pulsar timing arrays}",
    eprint = "1711.00075",
    archivePrefix = "arXiv",
    primaryClass = "astro-ph.HE",
    doi = "10.1093/mnras/sty689",
    journal = "Mon. Not. Roy. Astron. Soc.",
    volume = "477",
    number = "1",
    pages = "964--976",
    year = "2018"
}

@article{Reines:2015nyy,
    author = "Reines, Amy E. and Volonteri, Marta",
    title = "{Relations Between Central Black Hole Mass and Total Galaxy Stellar Mass in the Local Universe}",
    eprint = "1508.06274",
    archivePrefix = "arXiv",
    primaryClass = "astro-ph.GA",
    doi = "10.1088/0004-637X/813/2/82",
    journal = "Astrophys. J.",
    volume = "813",
    number = "2",
    pages = "82",
    year = "2015"
}

@article{Taylor:2021yjx,
    author = "Taylor, Stephen R.",
    title = "{The Nanohertz Gravitational Wave Astronomer}",
    eprint = "2105.13270",
    archivePrefix = "arXiv",
    primaryClass = "astro-ph.HE",
    month = "5",
    journal = "arXiv preprint",
    year = "2021"
}

@article{Middleton:2015oda,
    author = "Middleton, Hannah and Del Pozzo, Walter and Farr, Will M. and Sesana, Alberto and Vecchio, Alberto",
    title = "{Astrophysical constraints on massive black hole binary evolution from Pulsar Timing Arrays}",
    eprint = "1507.00992",
    archivePrefix = "arXiv",
    primaryClass = "astro-ph.CO",
    doi = "10.1093/mnrasl/slv150",
    journal = "Mon. Not. Roy. Astron. Soc.",
    volume = "455",
    number = "1",
    pages = "L72--L76",
    year = "2016"
}

@article{Chen:2016kax,
    author = "Chen, Siyuan and Middleton, Hannah and Sesana, Alberto and Del Pozzo, Walter and Vecchio, Alberto",
    title = "{Probing the assembly history and dynamical evolution of massive black hole binaries with pulsar timing arrays}",
    eprint = "1612.02826",
    archivePrefix = "arXiv",
    primaryClass = "astro-ph.HE",
    doi = "10.1093/mnras/stx475",
    journal = "Mon. Not. Roy. Astron. Soc.",
    volume = "468",
    number = "1",
    pages = "404--417",
    year = "2017",
    note = "[Erratum: Mon.Not.Roy.Astron.Soc. 469, 2455--2456 (2017)]"
}

@article{NANOGrav:2023pdq,
    author = "Agazie, Gabriella and others",
    collaboration = "NANOGrav",
    title = "{The NANOGrav 15 yr Data Set: Bayesian Limits on Gravitational Waves from Individual Supermassive Black Hole Binaries}",
    eprint = "2306.16222",
    archivePrefix = "arXiv",
    primaryClass = "astro-ph.HE",
    doi = "10.3847/2041-8213/ace18a",
    journal = "Astrophys. J. Lett.",
    volume = "951",
    number = "2",
    pages = "L50",
    year = "2023"
}

@article{Xue:2024qtx,
    author = "Xue, Xiao and Pan, Zhen and Dai, Liang",
    title = "{Non-Gaussian statistics of nanohertz stochastic gravitational waves}",
    eprint = "2409.19516",
    archivePrefix = "arXiv",
    primaryClass = "astro-ph.CO",
    doi = "10.1103/PhysRevD.111.043022",
    journal = "Phys. Rev. D",
    volume = "111",
    number = "4",
    pages = "043022",
    year = "2025"
}

@article{Bond:1990iw,
    author = "Bond, J. R. and Cole, S. and Efstathiou, G. and Kaiser, Nick",
    title = "{Excursion set mass functions for hierarchical Gaussian fluctuations}",
    reportNumber = "CFPA-TH-90-015",
    doi = "10.1086/170520",
    journal = "Astrophys. J.",
    volume = "379",
    pages = "440",
    year = "1991"
}

@article{Reardon:2023gzh,
    author = "Reardon, Daniel J. and others",
    title = "{Search for an Isotropic Gravitational-wave Background with the Parkes Pulsar Timing Array}",
    eprint = "2306.16215",
    archivePrefix = "arXiv",
    primaryClass = "astro-ph.HE",
    doi = "10.3847/2041-8213/acdd02",
    journal = "Astrophys. J. Lett.",
    volume = "951",
    number = "1",
    pages = "L6",
    year = "2023"
}

@article{Hellings:1983fr,
    author = "Hellings, R. W. and Downs, G. S.",
    title = "{Upper limits on the isotropic gravitational radiation background from pulsar timing analysis}",
    doi = "10.1086/183954",
    journal = "Astrophys.~J.~Lett.",
    volume = "265",
    pages = "L39--L42",
    year = "1983"
}

@article{Peters:1963ux,
    author = "Peters, P. C. and Mathews, J.",
    title = "{Gravitational radiation from point masses in a Keplerian orbit}",
    doi = "10.1103/PhysRev.131.435",
    journal = "Phys. Rev.",
    volume = "131",
    pages = "435--439",
    year = "1963"
}

@article{Sesana:2013wja,
    author = "Sesana, A.",
    title = "{Insights into the astrophysics of supermassive black hole binaries from pulsar timing observations}",
    eprint = "1307.2600",
    archivePrefix = "arXiv",
    primaryClass = "astro-ph.CO",
    doi = "10.1088/0264-9381/30/22/224014",
    journal = "Class. Quant. Grav.",
    volume = "30",
    pages = "224014",
    year = "2013"
}

@article{Merritt:2013awa,
    author = "Merritt, David",
    title = "{Loss-cone Dynamics}",
    eprint = "1307.3268",
    archivePrefix = "arXiv",
    primaryClass = "astro-ph.GA",
    doi = "10.1088/0264-9381/30/24/244005",
    journal = "Class. Quant. Grav.",
    volume = "30",
    pages = "244005",
    year = "2013"
}

@article{Tang:2017eiz,
    author = "Tang, Yike and MacFadyen, Andrew and Haiman, Zoltan",
    title = "{On the orbital evolution of supermassive black hole binaries with circumbinary accretion discs}",
    eprint = "1703.03913",
    archivePrefix = "arXiv",
    primaryClass = "astro-ph.HE",
    doi = "10.1093/mnras/stx1130",
    journal = "Mon. Not. Roy. Astron. Soc.",
    volume = "469",
    number = "4",
    pages = "4258--4267",
    year = "2017"
}

@article{Press:1973iz,
    author = "Press, William H. and Schechter, Paul",
    title = "{Formation of galaxies and clusters of galaxies by selfsimilar gravitational condensation}",
    doi = "10.1086/152650",
    journal = "Astrophys. J.",
    volume = "187",
    pages = "425--438",
    year = "1974"
}

@article{EPTA:2023fyk,
    author = "Antoniadis, J. and others",
    collaboration = "EPTA, InPTA:",
    title = "{The second data release from the European Pulsar Timing Array - III. Search for gravitational wave signals}",
    eprint = "2306.16214",
    archivePrefix = "arXiv",
    primaryClass = "astro-ph.HE",
    doi = "10.1051/0004-6361/202346844",
    journal = "Astron. Astrophys.",
    volume = "678",
    pages = "A50",
    year = "2023"
}

@article{Ellis:2023dgf,
    author = {Ellis, John and Fairbairn, Malcolm and H\"utsi, Gert and Raidal, Juhan and Urrutia, Juan and Vaskonen, Ville and Veerm\"ae, Hardi},
    title = "{Gravitational waves from supermassive black hole binaries in light of the NANOGrav 15-year data}",
    eprint = "2306.17021",
    archivePrefix = "arXiv",
    primaryClass = "astro-ph.CO",
    reportNumber = "KCL-PH-TH/2023-37, CERN-TH-2023-120, AION-REPORT/2023-06",
    doi = "10.1103/PhysRevD.109.L021302",
    journal = "Phys. Rev. D",
    volume = "109",
    number = "2",
    pages = "L021302",
    year = "2024"
}

@article{Girelli:2020goz,
    author = "Girelli, Giacomo and Pozzetti, Lucia and Bolzonella, Micol and Giocoli, Carlo and Marulli, Federico and Baldi, Marco",
    title = "{The stellar-to-halo mass relation over the past 12 Gyr: I. Standard $\Lambda$CDM model}",
    eprint = "2001.02230",
    archivePrefix = "arXiv",
    primaryClass = "astro-ph.CO",
    doi = "10.1051/0004-6361/201936329",
    journal = "Astron. Astrophys.",
    volume = "634",
    pages = "A135",
    year = "2020"
}

@article{NANOGrav:2023gor,
    author = "Agazie, Gabriella and others",
    collaboration = "NANOGrav",
    title = "{The NANOGrav 15 yr Data Set: Evidence for a Gravitational-wave Background}",
    eprint = "2306.16213",
    archivePrefix = "arXiv",
    primaryClass = "astro-ph.HE",
    doi = "10.3847/2041-8213/acdac6",
    journal = "Astrophys. J. Lett.",
    volume = "951",
    number = "1",
    pages = "L8",
    year = "2023"
}

@article{Phinney:2001di,
    author = "Phinney, E. S.",
    title = "{A Practical theorem on gravitational wave backgrounds}",
    eprint = "astro-ph/0108028",
    archivePrefix = "arXiv",
    journal = "arXiv preprint",
    month = "7",
    year = "2001"
}

@article{Cecchini:2026cru,
    author = "Cecchini, Chiara and El Gammal, Jonas and Franciolini, Gabriele and Pieroni, Mauro",
    title = "{Are PTA measurements sensitive to gravitational wave non-Gaussianities?}",
    eprint = "2605.05157",
    archivePrefix = "arXiv",
    primaryClass = "astro-ph.CO",
    reportNumber = "CERN-TH-2025-241",
    journal = "arXiv preprint",
    month = "5",
    year = "2026"
}

@article{Shen:2010aa,
    author = "Shen, Yue and others",
    title = "{A Catalog of Quasar Properties from SDSS DR7}",
    eprint = "1006.5178",
    archivePrefix = "arXiv",
    primaryClass = "astro-ph.CO",
    doi = "10.1088/0067-0049/194/2/45",
    journal = "Astrophys. J. Suppl.",
    volume = "194",
    pages = "45",
    year = "2011"
}

@article{Graham:2015tba,
    author = "Graham, Matthew J. and Djorgovski, S. G. and Stern, Daniel and Drake, Andrew J. and Mahabal, Ashish A. and Donalek, Ciro and Glikman, Eilat and Larsen, Steve and Christensen, Eric",
    title = "{A systematic search for close supermassive black hole binaries in the Catalina Real-Time Transient Survey}",
    eprint = "1507.07603",
    archivePrefix = "arXiv",
    primaryClass = "astro-ph.GA",
    doi = "10.1093/mnras/stv1726",
    journal = "Mon. Not. Roy. Astron. Soc.",
    volume = "453",
    number = "2",
    pages = "1562--1576",
    year = "2015"
}

@article{Xu:2023wog,
    author = "Xu, Heng and others",
    title = "{Searching for the Nano-Hertz Stochastic Gravitational Wave Background with the Chinese Pulsar Timing Array Data Release I}",
    eprint = "2306.16216",
    archivePrefix = "arXiv",
    primaryClass = "astro-ph.HE",
    doi = "10.1088/1674-4527/acdfa5",
    journal = "Res. Astron. Astrophys.",
    volume = "23",
    number = "7",
    pages = "075024",
    year = "2023"
}

@article{EPTA:2023xxk,
    author = "Antoniadis, J. and others",
    collaboration = "EPTA, InPTA",
    title = "{The second data release from the European Pulsar Timing Array - IV. Implications for massive black holes, dark matter, and the early Universe}",
    eprint = "2306.16227",
    archivePrefix = "arXiv",
    primaryClass = "astro-ph.CO",
    doi = "10.1051/0004-6361/202347433",
    journal = "Astron. Astrophys.",
    volume = "685",
    pages = "A94",
    year = "2024"
}

@article{Armitage:2002uu,
    author = "Armitage, Philip J. and Natarajan, Priyamvada",
    title = "{Accretion during the merger of supermassive black holes}",
    eprint = "astro-ph/0201318",
    archivePrefix = "arXiv",
    doi = "10.1086/339770",
    journal = "Astrophys. J. Lett.",
    volume = "567",
    pages = "L9--L12",
    year = "2002"
}

@article{Begelman:1980vb,
    author = "Begelman, M. C. and Blandford, R. D. and Rees, M. J.",
    title = "{Massive black hole binaries in active galactic nuclei}",
    doi = "10.1038/287307a0",
    journal = "Nature",
    volume = "287",
    pages = "307--309",
    year = "1980"
}

@ARTICLE{1993MNRAS.262..627L,
       author = {{Lacey}, Cedric and {Cole}, Shaun},
        title = "{Merger rates in hierarchical models of galaxy formation}",
      journal = {\mnras},
         year = 1993,
        month = jun,
       volume = {262},
       number = {3},
        pages = {627-649},
          doi = {10.1093/mnras/262.3.627},
       adsurl = {https://ui.adsabs.harvard.edu/abs/1993MNRAS.262..627L}
}

\end{document}